\documentclass{siamart171218}
\usepackage{amssymb}
\usepackage{amsmath}
\usepackage{mathtools}

\usepackage{setspace}
\usepackage{titlesec}
\usepackage{lmodern}
\usepackage{algpseudocode}
\newcommand{\review}[1]{\textcolor{black}{#1}}
\newcommand{\eqreview}[1]{\color{black}{#1}}
\newcommand{\makeblack}[1]{\color{black}{#1}}

\titleformat{\paragraph}
{\normalfont\normalsize\bfseries}{\theparagraph}{1em}{}
\titlespacing*{\paragraph}
{0pt}{3.25ex plus 1ex minus .2ex}{1.5ex plus 66.2ex}

\title{Displacement of transport processes on networked topologies}

\author{Daniel B. Wilson\thanks{Mathematical Institute, University of Oxford, Andrew Wiles Building, Woodstock Road, Oxford OX2 6GG, United Kingdom}
 \and Ruth E. Baker\footnotemark[1]
 \and Francis G. Woodhouse\footnotemark[1]}
 
\begin{document}
\maketitle
\begin{abstract}
Consider a particle whose position evolves along the edges of a network. One definition for the displacement of a particle is the length of the shortest path on the network between the current and initial positions of the particle. Such a definition fails to incorporate information of the actual path the particle traversed. In this work we consider another definition for the displacement of a particle on networked topologies. Using this definition, which we term the winding distance, we demonstrate that for Brownian particles, confinement to a network can induce a transition in the mean squared displacement from diffusive to ballistic behaviour, $\langle x^2(t) \rangle \propto t^2$ for long times. A multiple scales approach is used to derive a macroscopic evolution equation for the displacement of a particle and uncover a topological condition for whether this transition in the mean squared displacement will occur. Furthermore, for networks satisfying this topological condition, we identify a prediction of the timescale upon which the displacement transitions to long-time behaviour. Finally, we extend the investigation of displacement on networks to a class of anomalously diffusive transport processes, where we find that the mean squared displacement at long times is affected by both network topology and the character of the transport process.

\end{abstract}

\section{\label{sec:level1}Introduction}

The migration of stem cells during embryogenesis \cite{Aman2010}; the transport of proteins along microtubules within eukaryotic cells \cite{Welte2004}; and the rotational dynamics of polymers \cite{JungKim2015} are all examples of biological transport processes that take place in complex environments. Transport processes in such environments are not just of interest within the biological sciences, but are ubiquitous in the study of traffic flow \cite{Junevicius2009,Asaithambi2018} and human crowd management \cite{Hughes2002}. Many of these transport processes are inherently random, and they can be investigated through studying statistics of stochastic transport models. A critical transport statistic used to classify the character of a transport process is the mean squared displacement (MSD), which at time $t$ is denoted by $\langle x^2 (t) \rangle$. A particle whose position undergoes Brownian motion and lies on the real line, has MSD that evolves linearly in time, $\langle x^2(t) \rangle \propto t$. This is a defining feature of what is referred to as classical diffusion. For more general transport processes the MSD follows a power law, $\langle x^2 (t) \rangle \propto t^{\alpha}$, and the transport process is referred to as: sub-diffusive if $0 < \alpha < 1$; diffusive if $\alpha=1$; super-diffusive or sub-ballistic if $1 < \alpha < 2$; and ballistic if $\alpha=2$. 

For simple geometries such as the real line it is unambiguous how to define displacement: setting the origin of the real line to be at the initial position of the particle the displacement is defined as the current position of the particle, $x(t)$. However, for more complex geometries there are several ways to measure displacement. Consider a Brownian particle on a comb lattice \cite{Redner_GTFP, Benichou_PRL_combs, Burioni_PRE03, Illien_JPA16, Agliari_PRE16}, a topological structure consisting of an infinite backbone (the real line) with teeth (dendrites) of infinite length that protrude periodically out of the backbone. The MSD, when displacement is measured as the distance traversed along the backbone only, is sub-diffusive with $\langle x^2(t) \rangle \propto t^{1/2}$ for long times \cite{Illien_JPA16}. The particle spends large amounts of time diffusing along the protruding teeth, during which no transport along the backbone occurs. These long periods of time, where the displacement along the backbone does not change, are responsible for the sub-diffusive nature of the MSD. However, if we define displacement to also include the distance travelled down the teeth then we do not retain the sub-diffusive MSD. The choice of definition for displacement can therefore have a large impact on what we can learn about the transport process.

For topological structures that are embedded within a two- or three-dimensional domain a common definition of displacement is the Euclidean distance between the current and initial positions. With this definition, the potential for geometry to induce a qualitative change in the displacement of a particle has been widely investigated for a class of complex structures known as fractals \cite{Bhattacharya_PRE14, Guyer_PRA_84, Schulzky_SIGSAM_00, Dasgupta_JPA_99, Kozak_CPL_97, Huang_EPJ_06}. In fact, for Brownian particles on these structures, a random walk dimension, $d_w$, can be defined such that $\langle x^2(t) \rangle \propto t^{2/d_w}$ \cite{Gefen_PRL_80, Haber_JPA14}. However, if the topological structure cannot be embedded within a Euclidean domain then this measure of displacement is no longer appropriate. For random walks on general networked topologies (not necessarily embedded in a Euclidean domain), a measure of displacement sometimes used in the literature is the shortest path between the current and initial positions \cite{Gallos_PRE_04}. However, the shortest path does not include information about the actual path the particle took. Statistics of the path of a particle are of interest across research fields such as polymer physics \cite{RudnickHu_Polymer}, flux lines within superconductors \cite{Nelson_PRL}, and more broadly in the winding statistics of stochastic processes. Previous studies of winding statistics have focussed on the winding angle of a Brownian particle in a planar domain \cite{Spitzer1958,Comtet1990}. However, there has been little research on winding statistics for particles constrained to more complex environments. Kundu et. al \cite{Kundu_WindingStats} have studied winding statistics of a Brownian particle constrained to the unit circle. They calculate the exact distribution of the net number of clockwise or anticlockwise rotations as a function of time. We investigate the winding statistics of  particles constrained to more complex networks. We introduce a new definition of displacement, which we term the \textit{winding distance}, and investigate how it evolves over time as a function of network topology.



The remainder of this paper is organised as follows. In \Cref{sec:displacement_tree} we motivate and define our notion of displacement of a particle on a finite network by introducing a topological structure we term the \textit{displacement tree}. In order to analyse the displacement of a Brownian particle on a finite network at long times, we identify a relationship between displacement along the displacement tree and displacement on the half line with periodic bias. We calculate the strength of this bias as a function of the average degree of the network, a very simple topological property. In \Cref{PDEs} we study the displacement of a diffusive particle on the half line subject to periodic bias via a multiple scales approach introduced in \cite{ChapmanShabala17}. We derive a macroscopic advection-diffusion partial differential equation (PDE) to describe the evolution of the distribution of the winding distance of a particle. The diffusive and advective transport coefficients are calculated in terms of topological properties of the network. We uncover a topological condition that enables us to distinguish whether the long-time behaviour of the displacement of a Brownian particle is characteristically diffusive or ballistic, and a prediction of the timescale on which the transition between the two behaviours occurs. In \Cref{FracPDES} we generalise the transport process of the particle from Brownian motion to a continuous time random walk (CTRW) model of anomalous diffusion. Repeating the multiple scales approach in \Cref{PDEs} we derive a macroscopic fractional advection-diffusion PDE for the evolution of the winding distance of an anomaloulsy diffusive particle on a network. As in \Cref{PDEs}, we obtain a prediction of the timescale on which a particle transitions from the short-time to the long-time behaviour and discuss its asymptotic properties for small values of $\alpha$. Finally, in \Cref{sec:summary} we summarise our results and discuss the scope for future work.

\section{\label{sec:displacement_tree} Displacement on networks}

Consider a network $ \mathcal{G} = \{ \mathcal{V}, \mathcal{E} \} $,  where $\mathcal{V}$ is the set of vertices and $\mathcal{E}$ the set of edges. Each edge has an associated edge length $L_e > 0$ for all $e \in \mathcal{E}$. A point particle traverses the edges of the network, and on any given edge its position at time $t$ is denoted by $X(t)$. In order for the position $X(t)$ on the edge $e$ to uniquely identify the location of the particle on the network, each edge is oriented such that the vertex located at one end of the edge has position $0$ and the other vertex has position $L_e$ (the choice of which vertex has which position is arbitrary). The position of a particle on each edge evolves according to a standard Brownian motion. When a particle reaches a vertex, located at positions $0$ or $L_e$, an adjacent edge $e_{*}$ is selected uniformly at random\footnote{The standard numerical scheme for simulating a Brownian motion is the Euler-Maruyama method. For this scheme a particle performs jumps at discrete time steps of length $\Delta t$ and, as such, a particle will never reach a vertex but jump past it. For a discussion of how to implement the selection of adjacent edges see \Cref{app:simulations}.}\footnote{\review{The sampling procedure for the adjacent edge is formally incorporated in the definition of a Walsh Brownian motion, for more details see \cite{WalshBM}.}}. The new position of the particle is then either $X(t) = 0$ or $X(t) = L_{e_{*}}$ depending on how the edge $e_*$ is oriented. The initial condition for the particle can be either a position $x_0 \in (0,L_e)$ on a known edge $e$ of the network, or the particle can reside initially on a vertex of the network. In the latter case an initial edge is sampled uniformly from all adjacent edges to the initial vertex. 

\begin{figure}[tb]\label{Fig2.1}
\centering
$\begin{array}{c}
\includegraphics[scale=0.71]{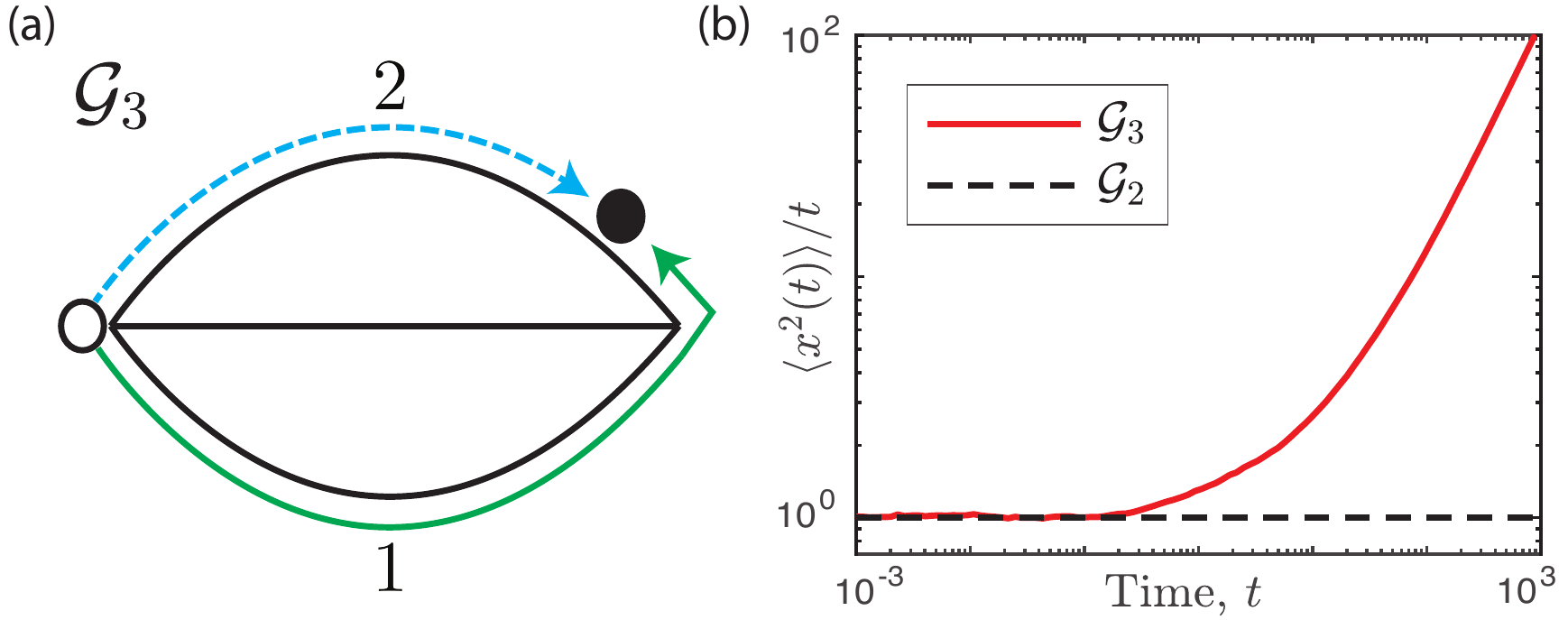}
\end{array}$
\caption{(a): Demonstrating the definition of displacement of a particle as the winding distance. The network $\mathcal{G}_3$ with two different trajectories that have the particle positioned at the same location on the network. (b): MSD for a particle on $\mathcal{G}_3$ calculated from $10^4$ realisations of a Brownian motion. The diffusion coefficient is $D=1/2$, the time-step is $\Delta t = 10^{-5}$ and the edge lengths are all equal to one.}
\end{figure}

Rather than the position of a particle at time $t$, it is often of interest to consider some measure of how far the particle has travelled up to and including time $t$. Such measures can be considered to be different definitions for the displacement of the particle. To aid discussion of different definitions of displacement, we introduce a simple class of networks. Let $\mathcal{G}_n$ be the network comprising two vertices connected by $n$ edges of the same length, $L$. The shortest path definition for displacement on networks sometimes used in the literature \cite{Gallos_PRE_04} is the length of the shortest path between the position at time $t$ and the initial position (at time $t=0$). In certain situations, for example when considering the rotation of polymers around cylindrical fibres \cite{RudnickHu_Polymer}, we are interested in the winding statistics of stochastic processes in constrained topologies \cite{Kundu_WindingStats}. Therefore, using the shortest path definition for displacement can be an undesirable transport statistic as it does not include any information about the winding of a particle between the initial and current times. Instead, consider a particle attached to an ``extendable leash''. In \Cref{Fig2.1}(a) we present two trajectories on the network $\mathcal{G}_3$. The initial position of the particle is the left-most vertex (\cref{Fig2.1}(a); unfilled circle), and one end of the leash remains stationary at this vertex at all times. The position of the particle evolves according to a Brownian motion. After selecting an initial edge uniformly at random, the particle moves around the network and the leash extends and contracts accordingly. To emphasise the distinction between the two definitions of displacement, in \Cref{Fig2.1}(a) we consider again the two possible trajectories. The first has the final position of the particle on the upmost edge with the leash hooked around the right-most vertex (\Cref{Fig2.1}(a); solid line with label one). The second trajectory has the same final position as the first. However, in this case, the particle travelled directly along the upmost edge (\Cref{Fig2.1}(a); dashed line with label two) and consequently the leash corresponding to the second trajectory is shorter. The length of the leash we refer to as the \textit{winding distance} of the particle and is the definition of displacement we consider in this work.

For the circle network, $\mathcal{G}_2$, the winding distance of a particle is equivalent to the displacement of a particle whose position evolves according to a Brownian motion on the infinite one-dimensional line \cite{Kundu_WindingStats, Redner_StatPhys}. Therefore, the winding distance, $x(t)$, of a Brownian particle at time $t$ on the unit circle is well studied, and the MSD is $\langle x^2(t) \rangle = 2D t$ for all times $t$, where $D$ is the diffusion coefficient. To investigate the evolution of the winding distance of a particle on the network $\mathcal{G}_3$  (\Cref{Fig2.1}(a)), we perform numerical realisations of a Brownian motion. From $10^4$ sample paths an estimate for the MSD is calculated and \Cref{Fig2.1}(b) shows the evolution of $\langle x^2(t) \rangle /t$ on a logarithmic scale. Initially $\langle x^2(t) \rangle = 2Dt$, as on short timescales the particle has not left the initial edge and so is classically diffusive. However as time increases, the solid (red) curve in \Cref{Fig2.1}(b) deviates away from the line $\langle x^2(t) \rangle = 2Dt$ (\Cref{Fig2.1}(b); dashed line) and in the long-time limit settles to $\langle x^2(t) \rangle \propto t^2$. This characteristic change in the MSD from diffusive to ballistic behaviour can be explained by considering the behaviour of a particle at the vertices. The first trajectory in \Cref{Fig2.1}(a) (solid line) reaches the right-most vertex via the bottom edge. At this vertex the particle then selects a new edge uniformly at random. If the new edge is either the top or middle edge (which occurs with total probability $2/3$), then moving along the new edge towards the left-most vertex will increase the displacement. However, if the new edge is the bottom edge (which occurs with probability $1/3$), then the displacement will decrease if the particle travels (back) along the bottom edge. 
\review{Consider temporarily, a particle which hops directly between the vertices in $\mathcal{G}_3$ over an edge, selected uniformly at random. The winding distance is now a stochastic process, which increases by $L$ if the new edge that the particle hops over is distinct from the previous edge, or decreases by $L$ if the new edge is the same as the previous edge. Thus the winding distance evolves as a biased random walk on $\{0,L,2L,\ldots\}$ with probability $2/3$ to jump in the positive direction. This is a well-studied random walk that is known to have a ballisitic MSD for all times $t$. Thus, the introduction of continuous stochastic motion along the edges of the network is responsible for the transition from diffusive to ballistic behaviour.}

For diffusive transport processes, can we identify which topologies will cause the displacement of a particle to become characteristically ballistic? Given a topology that will induce ballistic behaviour, can we predict the timescale on which this transition will occur and how it depends on properties of the network? How does topology affect transport statistics for other transport processes? In order to investigate these questions we first introduce a useful topological structure. 

\subsection{The displacement tree}

\begin{figure}[tb]\label{Fig2.2}
\centering
$\begin{array}{c}
\includegraphics[scale=0.61]{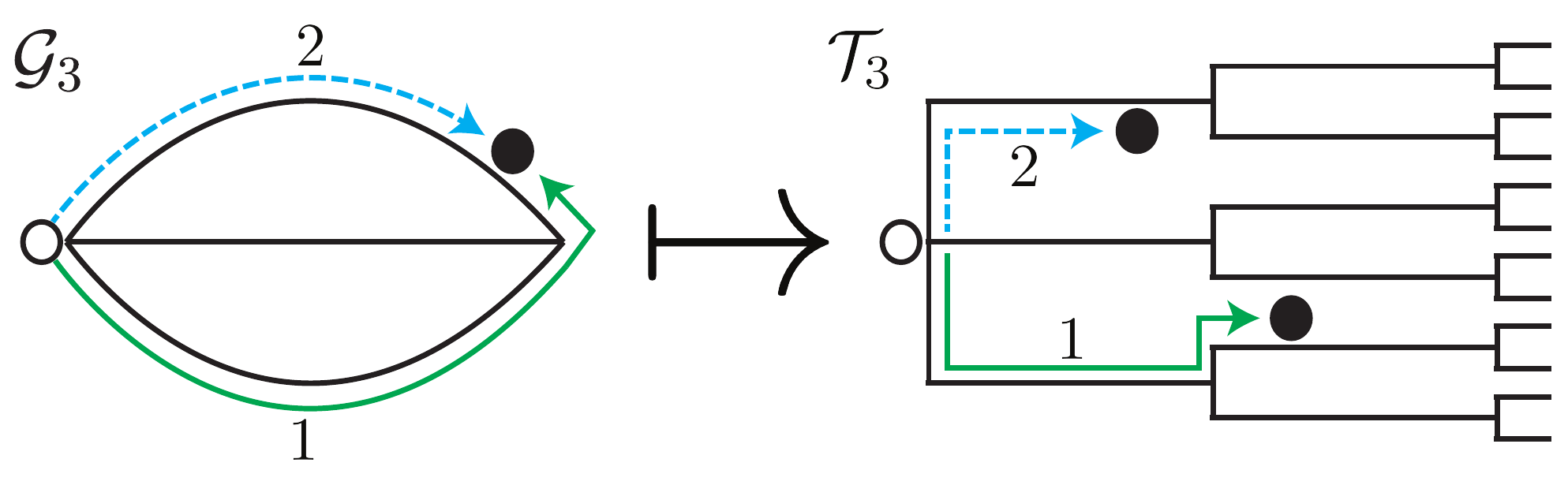}
\end{array}$
\caption{Demonstrating the equivalence of the winding distance of a particle on the network $\mathcal{G}_3$ and the displacement of a particle on the displacement tree $\mathcal{T}_3$. Two possible  trajectories that end at the same location on the network $\mathcal{G}_3$ are highlighted in dashed and dot-dashed lines. On the corresponding displacement tree, $\mathcal{T}_3$, the two trajectories have different displacements.}
\end{figure}

Recall that, for the circle $\mathcal{G}_2$, the winding distance of a particle is equivalent to the displacement of a particle on the real line. A similar relationship can be identified between the winding distance of a particle on a network and the displacement of a particle on a new topological structure we term the \textit{displacement tree}. To construct a displacement tree, $\mathcal{T}_3$, for the network $\mathcal{G}_3$ we need to select an initial position for the particle, which we take as the left-most vertex (see \Cref{Fig2.2}; unfilled circles). The three choices for the initial edge result in three potential initial trajectories for the particle. Consequently the tree $\mathcal{T}_3$ starts with its root, corresponding to the initial vertex, from which three parallel trajectories protrude (see \Cref{Fig2.2}). Now consider a particle that has just reached the right-most vertex of $\mathcal{G}_3$; the particle must choose a new edge. If it chooses the edge it just traversed and travels back along that edge the displacement of the particle will decrease. Travelling along the other two edges will result in an increase in displacement. To represent these possibilities, each of the three initial trajectories of the displacement tree branch into two new trajectories for the two edge choices that if travelled along will increase displacement. Repeating this procedure produces an infinite tree that spans the entire ensemble of possible trajectories for a particle on $\mathcal{G}_3$, oriented such that trajectories that increase displacement lead to the right. The winding distance of a Brownian particle on $\mathcal{G}_3$ is therefore equivalent to how far a Brownian particle has travelled along $\mathcal{T}_3$ (\Cref{Fig2.2}). 

The displacement tree $\mathcal{T}_3$ has a very simple structure: as the edges of $\mathcal{G}_3$ are all of equal length, $L$, the branching points appear on the tree periodically along all trajectories at positions $kL$ for integers $k \geq 1$. We define a protruding edge on the displacement tree to be an edge that emerges from a branching point and leads to the right. Both the vertices in $\mathcal{G}_3$ have degree three and so the number of protruding edges from any of the branching points in the displacement tree is the same, and equal to two. We refer to these displacement trees as regular: every possible trajectory of a particle on $\mathcal{T}_3$ of some fixed displacement will have passed the same number of branching points, all of which had the same number of protruding edges. As a result, we can equate the displacement of a particle on such a regular displacement tree with the displacement of a particle undergoing a new transport process on the non-negative half line. 

Consider a particle whose position on the regular displacement tree, $X(t)$, lies within the interval $[(k-1)L,kL]$ for some positive integer $k$. The position of the particle evolves according to a Brownian motion unless the particle is positioned on one of the boundaries of the interval, $X(t)=kL$ or $X(t)=(k-1)L$. These boundary positions correspond to the branching points on the displacement tree and, as such, a particle moves to the right of these points with probability $2/3$ and to the left with probability $1/3$. It is equivalent to just consider a particle on the non-negative half line, where the particle moves between the intervals $[(k-1)L,kL]$ for integers $k \geq 1$. For a particle with position $X(t)=kL$, the probability $2/3$ to move to the right is incorporated as a Robin boundary condition. The particle is absorbed into the new interval $[kL,(k+1)L]$ (a move to the right along the displacement tree) with probability $2/3$ and reflected back into the interval $[(k-1)L,kL]$ (a move to the left along the displacement tree) with probability $1/3$. For a particle positioned at $X(t)=(k-1)L$, the particle is absorbed into the new interval $[(k-2)L, (k-1)L]$ (a move to the left along the displacement tree) with probability $1/3$ and reflected back into the interval $[(k-1)L,kL]$ (a move to the right along the displacement tree) with probability $2/3$. This transport process is a Brownian motion on the non-negative half line with singular periodic bias of strength $2/3$.

\subsection{Extending to general networked topologies}

\begin{figure}[tb]\label{Fig2.3}
\centering
$\begin{array}{c}
\includegraphics[scale=0.45]{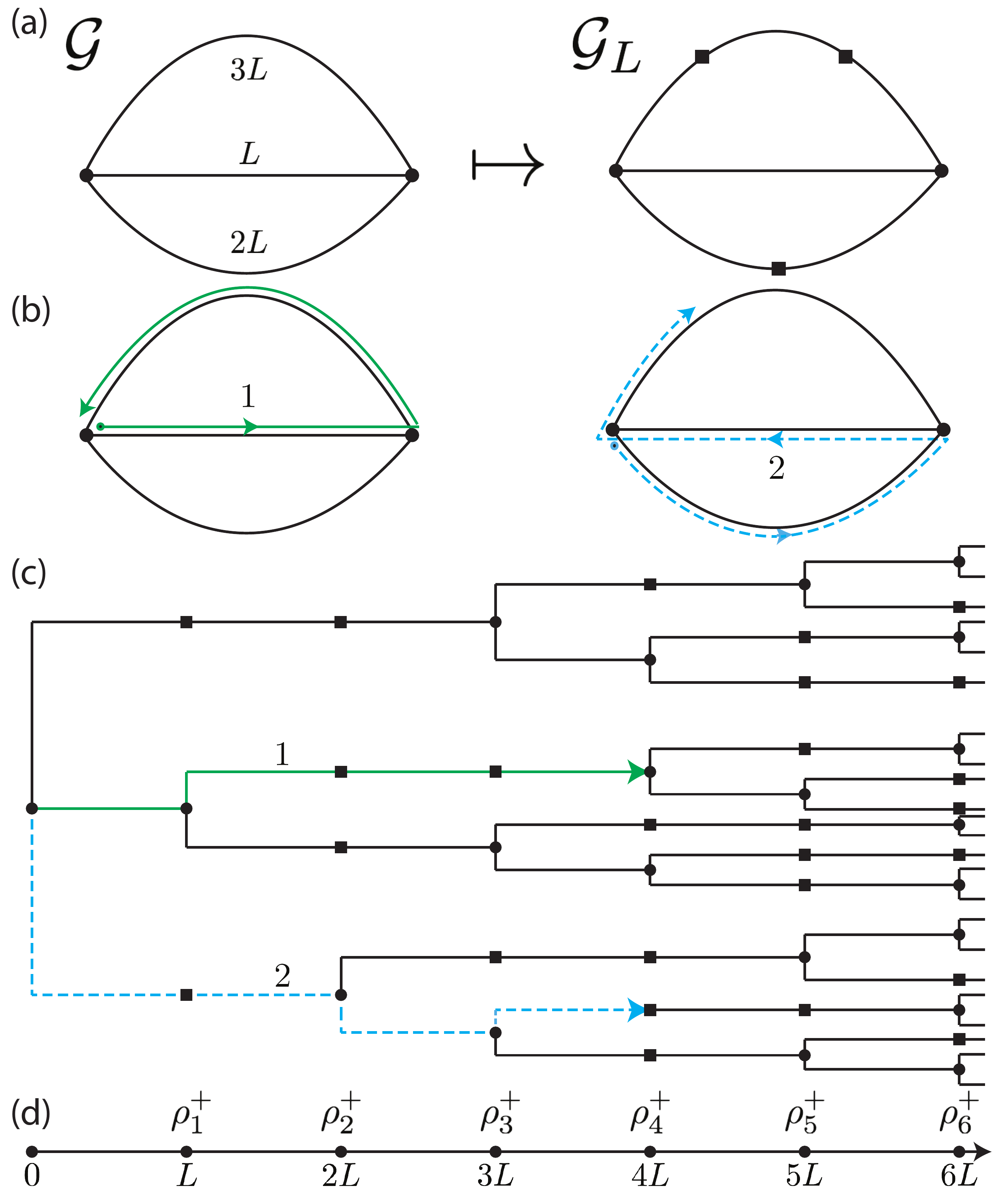}
\end{array}$
\caption{(a): Demonstrating the mapping of a network $\mathcal{G}$ to the topologically equivalent network $\mathcal{G}_L$ by introducing additional vertices here represented as squares. (b): Two possible trajectories each of length $4L$ are highlighted with solid and dashed lines on the network $\mathcal{G}$. The first trajectory ends at a vertex, whereas the second trajectory ends midway through an edge. (c): The asymmetric displacement tree for the network $\mathcal{G}$ and $\mathcal{G}_L$. The additional vertices introduced on the network $\mathcal{G}_L$ are represented as black squares on the tree. The two trajectories considered in (b) are highlighted on the tree using solid and dashed lines. (d): The half line $[0,\infty)$ with vertices at positions $kL$ for integers $k \geq 1$. The sequence of effective biases are shown $\{\rho_1^+,\rho_2^+,\ldots\}$.}
\end{figure}

Networks that represent complex environments will not necessarily have regular displacement trees. For a general network, $\mathcal{G} = \{\mathcal{V},\mathcal{E}\}$, the edge lengths, $L_e$ for $e \in \mathcal{E}$, are not always of equal length and the vertices will not always have equal degree. An example of such a network is shown in \Cref{Fig2.3}(a) where two vertices are connected by three multiple edges, similarly to $\mathcal{G}_3$, but now the top, middle and bottom edges have edge lengths $3L$, $L$ and $2L$, respectively. Consider two possible trajectories for a particle starting at the left-most node; trajectory one is highlighted with a solid line in \Cref{Fig2.3}(b), has a length of $4L$, and finishes at a vertex on the network $\mathcal{G}$, i.e. a branching point on the displacement tree (see \Cref{Fig2.3}(c); solid line with label one). Trajectory two is highlighted with a dashed line in \Cref{Fig2.3}, also has a length of $4L$ but finishes along an edge, not at a branching point. Therefore knowledge of the displacement of a particle is not sufficient to determine whether the particle is at a branching point on the displacement tree. In order to circumvent this problem we introduce a topologically equivalent network $\mathcal{G}_L = \{ \mathcal{V}_L, \mathcal{E}_L \}$ which is formed by introducing additional vertices into $\mathcal{G}$ that occur with equal-spacing of length $L$ (see \Cref{Fig2.3}(a); black squares). The corresponding displacement tree for $\mathcal{G}_L$ has branching points that occur periodically at positions $kL$ for integers $k \geq 1$ along every trajectory (see \Cref{Fig2.3}(c); black squares). Note that all the additional branching points introduced will have only one protruding edge as the additional vertices in $\mathcal{G}_L$ have degree two. As a result, they will not introduce a bias. For general networks $\mathcal{G} = \{ \mathcal{V},\mathcal{E}\}$, as long as the edge lengths, $L_e$, are rational, or can be approximated as such, there will always exist $L > 0$ such that $L_e = m_e L$ for some positive integer $m_e$ for all $e \in \mathcal{E}$, and so the transformation from $\mathcal{G}$ to $\mathcal{G}_L$ can be applied.

As for $\mathcal{G}_3$, for these more general networks $\mathcal{G}_L$, it would be convenient to equate the displacement of a particle diffusing on the displacement tree to the displacement of a particle on the half line. The displacement tree for $\mathcal{G}_L$ has branching points along all trajectories at positions $kL$ for integers $k \geq 1$. However, the number of protruding edges is not constant across trajectories with the same length. Consider again the highlighted trajectories of length $4L$ in \Cref{Fig2.3}(c). The branching point at the end of trajectory one has two protruding edges. A particle at this branching point travels along an edge that will increase displacement with probability $2/3$. However, trajectory two ends at a branching point with only one protruding edge. The probability of a particle travelling along an edge that will increase displacement is $1/2$. To make progress, we consider a particle on the displacement tree whose position lies in the interval $[(k-1)L,kL]$. For a particle at the boundary position $X(t)=kL$ we seek an averaged probability, $\rho^+_k$, to select an edge on the displacement tree which lies to the right of the branching points at $kL$. The probability of moving further along the displacement tree if the particle is at vertex $\nu \in \mathcal{V}_L$ is $(d_{\nu}-1)/d_{\nu}$  where $d_{\nu}$ is the degree of the vertex $\nu$. Weighting this by $p_k(\nu)$, the probability that the particle occupies vertex $\nu$ given it has displacement $kL$, and summing over $\nu \in \mathcal{V}_L$, yields
\begin{equation}\label{rho_k}
\rho^+_k = \sum_{\nu \in \mathcal{V}_L} \dfrac{d_{\nu}-1}{d_{\nu}} \times p_k(\nu).
\end{equation}
The probabilities, $\rho^+_k$, are used to define the Robin boundary conditions for the transport process on the half line, just as for symmetric displacement trees. Consider a particle in the interval $[(k-1)L,kL]$ with position at the boundary $X(t)=kL$. The particle is absorbed into the adjacent interval $[kL,(k+1)L]$ (a move to the right along the asymmetric displacement tree) with probability $\rho^+_k$. The particle is reflected back into the interval $[(k-1)L,kL]$ (a move to the left along the asymmetric displacement tree) with probability $1-\rho^+_k$. Therefore, \Cref{rho_k} provides the sequence of probabilities, $\{\rho^+_1, \rho^+_2, \ldots\}$, to be absorbed at position $kL$ into the interval $[kL,(k+1)L]$ (see \Cref{Fig2.3}(e)). The probabilities $p_k(\nu)$ are difficult to calculate as they depend heavily on the topology of the network and the initial position of the particle. However, the displacement of a particle transitions to ballistic behaviour when the periodic bias felt at positions $kL$ dominates over the unbiased Brownian motion. As such, we consider the displacement of a particle in the long-time limit, which is when the particle has traversed sufficiently many edges for the bias to dominate. The branching points in the displacement tree (corresponding to vertices in the network with degree greater than two) introduce an asymmetry, biasing particles further along the tree (to the right). As a particle moves further along the tree we are interested in the limit to which the sequence of effective biases $\{\rho^+_1, \rho^+_2, \ldots\}$ converges. Assume temporarily that the networks we consider are both finite and aperiodic. Then, as $k \rightarrow \infty$ the probability $p_k(\nu)$ converges to the equilibrium distribution for a discrete random walk on the vertices of $\mathcal{G}_L$ which is $p_{\infty}(\nu) = d_{\nu} / \sum_{\omega \in \mathcal{V}_L} d_{\omega}$ \footnote{For general finite Markov Chains an equilibrium distribution for the occupany probability of each state in the chain exists if and only if the chain (here the network) is irreducible and aperiodic. All connected undirected networks however are necessarily irreducible.}. The sequence $\{\rho^+_1, \rho^+_2, \ldots\}$ therefore converges to the limit
\begin{equation}\label{rho_inf}
\rho^+_{\infty} = \sum_{\nu \in \mathcal{V}_L} \dfrac{d_{\nu}-1}{d_{\nu}} \times \dfrac{d_{\nu}}{\sum_{\omega \in \mathcal{V}_L} d_{\omega}} = 1 - \dfrac{1}{\bar{d}},
\end{equation}
where $\bar{d} = 2|\mathcal{E}_L| / |\mathcal{V}_L|$ is the average degree of $\mathcal{G}_L$. \Cref{rho_inf} provides an approximation for the probability of a particle to be absorbed at $kL$ into the interval $[kL,(k+1)L]$ once the particle is sufficiently far along the half line, that is, for large integers $k$. Note that the assumption that the network is finite is not a necessary one. Some infinite networks, such as a honeycomb lattice, have displacement trees where the equilibrium distribution $p_{\infty}(\nu)$ does exist. We can also relax the assumption that the network is aperiodic; see \Cref{app:periodic}.

In summary, in the long-time limit we have identified a link between the displacement of a diffusing particle on a general network and displacement on the half line with a constant periodic bias. The focus of the next section is to use multiple scales analysis to explore the transport properties of the process on the half line.

\section{\label{PDEs} Periodic bias on the half line: multiple scales analysis}

In this section we consider the transport properties of a Brownian motion on the half line with periodic bias using \review{an approach from a recent paper by Chapman and Shabala \cite{ChapmanShabala17}. The paper introduces a method of multiple scales to derive the macroscopic transport equations of a random walk on a periodic lattice with spatially dependent transition rates. The slow scale is continuous and evolves as the particle walks over many periods of the lattice. The fast scale is discrete and is defined on a unit periodic interval made up of $N$ lattice sites. In contrast to \cite{ChapmanShabala17}, in this work we are interested in the transport of a Brownian particle and so we take the limit $N \rightarrow \infty$ to obtain a continuous expression for the fast scale as well as the slow scale.}

\begin{figure}[tb]\label{Fig3.1}
\centering
$\begin{array}{c}
\includegraphics[scale=0.65]{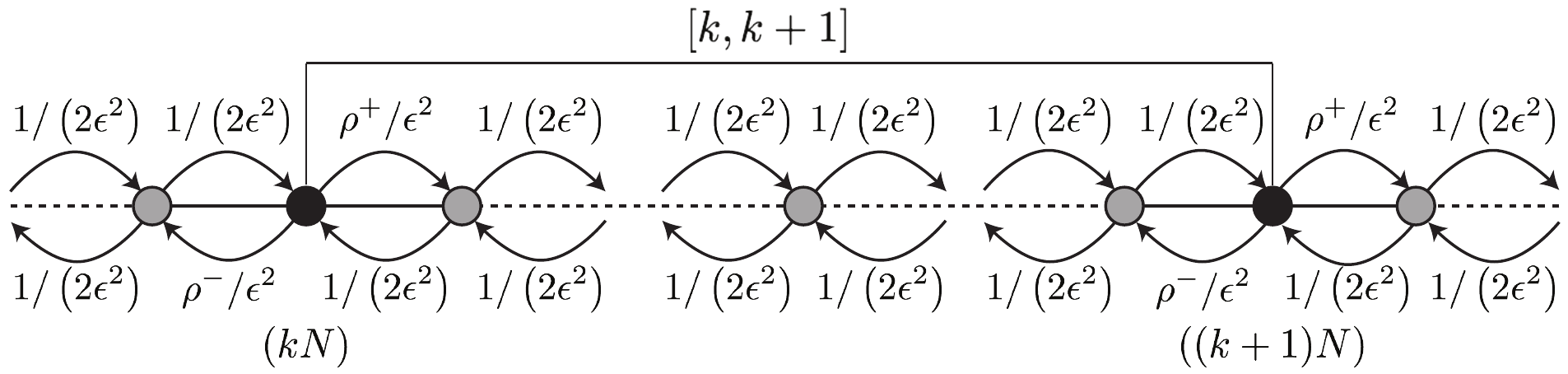}
\end{array}$
\caption{A diagrammatic representation of the CTMC on the discretised interval $[k,k+1]$. The black circles represent the lattice sites which correspond to the biased branching points, the grey circles are the lattice sites where symmetric transport occurs.}
\end{figure}

Brownian motion is defined in continuous space, however to make progress we formulate a continuous time Markov Chain (CTMC) in discrete space that has the same macroscopic properties. We non-dimensionalise in space by setting the length of each interval to be one ($L=1$), the intervals $[k,k+1]$ are then discretised into $N+1$ lattice sites each separated by distance $\epsilon = 1/N$ (see \cref{Fig3.1}). Consider the position of a particle on the discretised lattice that evolves according to a CTMC, where the transition rates to leave each site are symmetric and chosen to be $ \lambda_{n \rightarrow n-1} = \lambda_{n \rightarrow n+1} = D/\epsilon^2$. In the limit $\epsilon \rightarrow 0$ and $N \rightarrow \infty$, the distribution of the position of a particle which evolves according to this CTMC is equivalent to the same distribution arising from a Brownian motion on the half line with diffusion coefficient $D$. For the remainder of this work we take $D=1/2$. To include the periodic bias at branching points, which occur at lattice sites with indices $kN$ for all integers $k > 0$, we choose asymmetric transition rates at these points. For a particle to exit on the right of the branching point with probability $\rho^+_{\infty}$ we choose transition rates $\lambda_{kN \rightarrow kN+1} = \rho^+ / \epsilon^2$ and $\lambda_{kN \rightarrow kN-1} = \rho^- / \epsilon^2$, where $\rho^- = 1 - \rho^+$ and we have dropped the $\infty$ subscripts from \Cref{sec:displacement_tree}. For a diagrammatic representation of the CTMC see \Cref{Fig3.1}.

Let $p_n(t)$ be the probability a particle occupies lattice site $n \geq 0$ at time $t$. The master equation for $p_n(t)$ is
\begin{equation}\label{master}
\dfrac{\mathrm{d} p_n}{\mathrm{d} t} = \lambda_{(n-1)\rightarrow n} p_{n-1} + \lambda_{(n+1)\rightarrow n} p_{n+1} - \left( \lambda_{n \rightarrow (n-1)} + \lambda_{n \rightarrow (n+1)} \right) p_n.
\end{equation}
Initial conditions imply $p_0(0)=1$ and $p_i(0)=0$ for all $i \geq 1$ as the initial winding distance is always zero. The winding distance is defined such that it is always non-negative, therefore we impose a reflective boundary condition at the origin by setting $\lambda_{-1 \rightarrow 0} = \lambda_{0 \rightarrow -1} = 0$ and $\lambda_{0 \rightarrow 1} = 1/\epsilon^2$. We define the spatial coordinate $x= \epsilon n$, and note that, for $\epsilon \ll 1$, $x$ evolves on a slower time scale than the lattice index $n$. Following \cite{ChapmanShabala17}, let $p_n(t) = P_n(x,t)$ so the probability density for the position of a particle at time $t$ depends on both the fast scale, $n$, and the slow scale $x$. Re-writing \cref{master} in terms of $P_n(x,t)$, we have
\begin{eqnarray}
\dfrac{\partial P_n(x,t)}{\partial t} &=& \lambda_{(n-1)\rightarrow n} P_{n-1}\left(x-\epsilon,t \right) + \lambda_{(n+1)\rightarrow n} P_{n+1}\left(x+\epsilon,t \right)  \label{eq:P_nxt} \\
& & - \left( \lambda_{n \rightarrow (n-1)} + \lambda_{n \rightarrow (n+1)} \right) P_n\left(x,t \right), \notag
\end{eqnarray}
for $n \geq 0$ and $x \in [0,\infty)$. To derive a macroscopic PDE in the spatial variable $x$, we assume that the variables $x$ and $n$ are independent. This assumption reduces the infinite set of equations (\ref{eq:P_nxt}) to a finite set of $N$ distinct equations. Introducing circular notation $\overline{i \pm 1} = \left( {i \pm 1} - 1 \text{ mod } N \right) + 1$, we write the master equation for the probabilities $P_i(x,t)$ as
\begin{eqnarray}
\label{eq:mastereqn}
\dfrac{\partial P_i}{\partial t}(x,t) &=& \lambda_{\overline{i-1}\rightarrow i} P_{\overline{i-1}}(x-\epsilon,t) + \lambda_{\overline{i+1}\rightarrow i} P_{\overline{i+1}}(x+\epsilon,t) \\
&& - \left( \lambda_{i \rightarrow \overline{i+1}} + \lambda_{i \rightarrow \overline{i-1}}\right) P_i(x,t), \notag
\end{eqnarray}
for $i \in \{1,\ldots, N\}$ and $x \in [0,\infty)$. The finite set of equations defines the periodic unit interval. On the periodic unit interval there are symmetric transition rates of $1/\left(2 \epsilon^2\right)$ for all adjacent sites, other than the rates $\lambda_{N \rightarrow 1} = \rho^+ / \epsilon^2$ and $\lambda_{N \rightarrow N-1} = \rho^- /\epsilon^2$.


Following the derivation in \cite{ChapmanShabala17}, we introduce $\vec{P}(x,t) = \left[ P_1(x,t) , \ldots , P_N(x,t) \right]^T$ and expand about $x$ to obtain
\begin{equation}\label{eq:PDE_eps}
\epsilon^2 \dfrac{\partial \vec{P}}{\partial t}(x,t) = \mathbf{A} \vec{P} + \epsilon \mathbf{B} \dfrac{\partial \vec{P}}{\partial x} + \epsilon^2 \mathbf{C} \dfrac{\partial^2 \vec{P}}{\partial x^2} + \mathcal{O}\left( \epsilon^3 \right),
\end{equation}
where 
\begin{subequations}\label{eq:matrices}
\begin{equation}\label{eq:matricesA}
\eqreview{\mathbf{A} = \begin{pmatrix}
-1 & 1/2 & 0 & \cdots & 0 & 0 & \rho^+ \\
1/2 & -1 & 1/2 & \cdots & 0 & 0 & 0 \\
0 & 1/2 & -1 & \cdots & 0 & 0 & 0 \\
0 & 0 & 1/2 & \cdots & 0 & 0 & 0 \\
& \vdots & & \ddots & & \vdots & \\
0 & 0 & 0 & \cdots & 1/2 & 0 & 0 \\
0 & 0 & 0 & \cdots & -1 & 1/2 & 0 \\
0 & 0 & 0 & \cdots & 1/2 & -1 & \rho^- \\
1/2 & 0 & 0 & \cdots & 0 & 1/2 & -1
\end{pmatrix},}
\end{equation}
\begin{equation}
\eqreview{\mathbf{B} = \begin{pmatrix}
0 & 1/2 & 0 & \cdots & 0 & 0 & -\rho^+ \\
-1/2 & 0 & 1/2 & \cdots & 0 & 0 & 0 \\
0 & -1/2 & 0 & \cdots & 0 & 0 & 0 \\
0 & 0 & -1/2 & \cdots & 0 & 0 & 0 \\
& \vdots & & \ddots & & \vdots & \\
0 & 0 & 0 & \cdots & 1/2 & 0 & 0 \\
0 & 0 & 0 & \cdots & 0 & 1/2 & 0 \\
0 & 0 & 0 & \cdots & -1/2 & 0 & \rho^- \\
1/2 & 0 & 0 & \cdots & 0 & -1/2 & 0
\end{pmatrix},}
\end{equation}
\begin{equation}
\eqreview{\mathbf{C} = \begin{pmatrix}
0 & 1/4 & 0 & \cdots & 0 & 0 & \rho^+/2 \\
1/4 & 0 & 1/4 & \cdots & 0 & 0 & 0 \\
0 & 1/4 & 0 & \cdots & 0 & 0 & 0 \\
0 & 0 & 1/4 & \cdots & 0 & 0 & 0 \\
& \vdots & & \ddots & & \vdots & \\
0 & 0 & 0 & \cdots  & 1/4 & 0 & 0 \\
0 & 0 & 0 & \cdots & 0 & 1/4 & 0 \\
0 & 0 & 0 & \cdots & 1/4 & 0 & \rho^-/2 \\
1/4 & 0 & 0 & \cdots & 0 & 1/4 & 0
\end{pmatrix}.}
\end{equation}
\end{subequations}
\review{Note that when $\rho^+ = \rho^- = 1/2$ we have symmetric transition rates of $1/(2\epsilon^2)$ at all sites on the unit interval. These rates were chosen such that the time derivative on the right-hand side of \cref{eq:PDE_eps} appears at the same order in $\epsilon$ as the second order spatial derivatives on the left-hand side. Therefore, as $\epsilon \rightarrow 0$, the probability distribution for the displacement of a particle follows a macroscopic diffusion equation. However, when $\rho^+ \neq \rho^-$,} recall that the displacement of a particle on a networked topology may transition from diffusive to ballistic behaviour. As such, we anticipate both advective and diffusive contributions to the macroscopic PDE describing the displacement of a particle and we rescale time, $\hat{t} = \epsilon^{-1} t$, to give
\begin{equation}\label{eq:driftdom}
\epsilon \dfrac{\partial \vec{P}}{\partial \hat{t}}(x,\hat{t}) = \mathbf{A} \vec{P} + \epsilon \mathbf{B} \dfrac{\partial \vec{P}}{\partial x} + \epsilon^2 \mathbf{C} \dfrac{\partial^2 \vec{P}}{\partial x^2},
\end{equation}
\review{where the time-dependent derivative now appears at the same order of magnitude as the first order spatial derivative.} We make the series expansion $\vec{P}(x,\hat{t}) = \vec{P}_0(x,\hat{t}) + \epsilon \vec{P}_1(x,\hat{t}) + \epsilon^2 \vec{P}_2(x,\hat{t}) + \ldots$ and substitute into \cref{eq:driftdom}. Collecting terms of $\mathcal{O}(\epsilon^0)$ yields $\mathbf{A} \vec{P}_0 = \vec{0}$, which has solution $\vec{P}_0 = f(x,\hat{t} ) \eqreview{\vec{u}_0(\epsilon)}$ \review{where  $f(x,\hat{t})$ is some function to be determined and $\eqreview{\vec{u}_0}(\epsilon)$ is in the kernel of $\mathbf{A}$\footnote{The matrix $\mathbf{A}$ is the rate matrix for the CTMC on the periodic interval, therefore as $\vec{u}_0(\epsilon)$ is in the kernel of $\mathbf{A}$ it is proportional to the equilibrium distribution of the CTMC.} which is given by
\begin{equation}\label{eq:u0}
\left[ \vec{u}_{0}(\epsilon) \right]_n = 2 \rho^+ \epsilon - 2n \left( \rho^+ - \rho^- \right)\epsilon^2,
\end{equation}
for $1 \leq n \leq N-1$ and $\left[ \vec{u}_{0}(\epsilon) \right]_N = \epsilon$, such that $\sum_{n=1}^N \left[ \vec{u}_{0}(\epsilon)\right]_n=1$.} Collecting terms of $\mathcal{O}\left( \epsilon^1 \right)$ we find
\begin{equation}
\dfrac{\partial \vec{P}_0}{\partial \hat{t}} = \mathbf{A} \vec{P}_1 + \mathbf{B} \dfrac{\partial \vec{P}_0}{\partial x}.
\end{equation}
Substituting in the expression for $\vec{P}_0$ and rearranging yields
\begin{equation}\label{eq:order1}
\mathbf{A} \vec{P}_1 = \eqreview{\vec{u}_0(\epsilon)} \makeblack{ \dfrac{\partial f}{\partial \hat{t}}  - \mathbf{B}} \eqreview{\vec{u}_0(\epsilon)} \makeblack{\dfrac{\partial f}{\partial x}}.
\end{equation}
Applying the Fredholm alternative theorem\footnote{The Fredholm alternative theorem states that $\mathbf{M} \vec{x} = \vec{b}$ has a solution if and only if for all $\vec{y}$ such that $\mathbf{M}^T \vec{y}=\vec{0}$ we also have $\vec{y}^T \vec{b} = \vec{0}$.} we see that, in order for \cref{eq:order1} to have a solution, we need the right-hand side to be orthogonal to the null space of $\mathbf{A}^T$. The null space is spanned by the vector $\vec{v} =[1,\ldots,1]^T$, and the Fredholm alternative theorem therefore implies
\begin{equation}\label{eq:f}
\dfrac{\partial f}{\partial \hat{t}} = \dfrac{\vec{v}^T \mathbf{B} \eqreview{\vec{u}_0(\epsilon)}}{\vec{v}^T \eqreview{\vec{u}_0(\epsilon)}} \dfrac{\partial f}{\partial x}.
\end{equation} 
\review{Introducing $V(\epsilon)  = - \vec{v}^T \mathbf{B} \vec{u}_0(\epsilon) / \vec{v}^T \vec{u}_0(\epsilon) $ we can solve \cref{eq:order1} to give $\vec{P}_1 = f_x \vec{u}_1(\epsilon) + g \vec{u}_0(\epsilon)$, where $g(x,\hat{t})$ is some function to be determined, $\vec{u}_1(\epsilon)$ is such that $\mathbf{A}\vec{u}_1(\epsilon) = - \left( V (\epsilon)  \mathbf{I} + \mathbf{B} \right) \vec{u}_0(\epsilon)$ and $\mathbf{I}$ is the identity matrix. Noting that $V(\epsilon)  = \epsilon \left( \rho^+ - \rho^- \right)$ we can solve for $\vec{u}_1(\epsilon)$ to obtain
\begin{equation}
\left[ \vec{u}_{1}(\epsilon)\right]_n = \beta_0(\epsilon) + \beta_1(\epsilon) n + \beta_2(\epsilon) n^2 + \beta_3(\epsilon) n^3,
\end{equation}
for $1\leq n \leq N-1$ where the coefficients are given by
\begin{subequations}
\begin{eqnarray}
\beta_0(\epsilon) & = & 2\rho^+ \epsilon + \dfrac{\rho^+\left( \rho^+ - \rho^- \right)}{3} \left( 1 - \epsilon^2 \right), \\
\beta_1(\epsilon) & = & - \left(\rho^+ - \rho^-\right)\epsilon - 2\left( \rho^+ - \rho^- \right) \epsilon^2 + \dfrac{\left( \rho^+ - \rho^- \right)^2}{3}\epsilon^3, \\
\beta_2(\epsilon) &=& 2 \rho^- \left( \rho^+ - \rho^- \right)\epsilon^2, \\
\beta_3(\epsilon) &=& \dfrac{2\left( \rho^+ - \rho^- \right)^2}{3}\epsilon^3, 
\end{eqnarray}
\end{subequations}
and the final term in $\vec{u}_1(\epsilon)$ is given by
\begin{equation}\label{eq:u1}
\left[ \vec{u}_{1}(\epsilon)\right]_N = \epsilon + \dfrac{\rho^+ - \rho^-}{6}\left( 1-\epsilon^2\right).
\end{equation}} 
We now proceed to collect terms of $\mathcal{O}\left( \epsilon^2 \right)$ which gives
\begin{equation}\label{eq:order2}
\dfrac{\partial \vec{P}_1}{\partial \hat{t}} = \mathbf{A} \vec{P}_2 + \mathbf{B} \dfrac{\partial \vec{P}_1}{\partial x} + \mathbf{C} \dfrac{\partial^2 \vec{P}_0}{\partial x^2}.
\end{equation}
As before, we rearrange \cref{eq:order2} to give
\begin{equation}
\mathbf{A} \vec{P}_2 = \review{\vec{u}_0(\epsilon)}\dfrac{\partial g}{\partial \hat{t}}  -  \review{V(\epsilon) \vec{u}_1(\epsilon)}\dfrac{\partial^2 f}{\partial x^2} - \mathbf{B}\review{\vec{u}_1(\epsilon)} \dfrac{\partial^2 f}{\partial x^2} - \mathbf{C}\review{\vec{u}_0(\epsilon)}\dfrac{\partial^2 f}{\partial x^2} - \mathbf{B}\review{\vec{u}_0(\epsilon)} \dfrac{\partial g}{\partial x}.
\end{equation}
Applying the Fredholm alternative theorem once again implies
\begin{equation}\label{eq:g}
\dfrac{\partial g}{\partial \hat{t}} + \review{V(\epsilon) } \dfrac{\partial g}{\partial x} = \dfrac{\vec{v}^T \mathbf{B}\review{\vec{u}_1(\epsilon)} + \vec{v}^T \mathbf{C}\review{\vec{u}_0(\epsilon)} + \review{V(\epsilon)} \vec{v}^T\review{\vec{u}_1(\epsilon)}}{\vec{v}^T\review{\vec{u}_0(\epsilon)}}~\dfrac{\partial^2 f}{\partial x^2}.
\end{equation}
We can now rewrite the power series as
\begin{equation}\label{eq:expansion}
\vec{P} = \vec{P}_0 + \epsilon \vec{P}_1 + \ldots = (f(x,\hat{t}) + \epsilon g(x,\hat{t})) \review{\vec{u}_0(\epsilon) + \epsilon f_x(x,\hat{t}) \vec{u}_1(\epsilon)} + \ldots.
\end{equation}
Introducing \review{$h_{\epsilon}(x,\hat{t}) = f(x,\hat{t}) + \epsilon g(x,\hat{t})$}, rescaling into time coordinates $t = \epsilon \hat{t}$ and using \cref{eq:f} and \cref{eq:g} we have
\begin{equation}\label{eq:h_epsilon}
\review{\epsilon \dfrac{\partial h_{\epsilon}}{\partial t} + V (\epsilon) \dfrac{\partial h_{\epsilon}}{\partial x} = \epsilon D (\epsilon ) \left( \dfrac{\partial^2 h_{\epsilon}}{\partial x^2} - \epsilon \dfrac{\partial^2 g}{\partial x^2}\right),}
\end{equation} 
where $\review{D(\epsilon)  = \left( \vec{v}^T \mathbf{B}\vec{u}_1(\epsilon) + \vec{v}^T \mathbf{C}\vec{u}_0(\epsilon) + V(\epsilon) \vec{v}^T\vec{u}_1(\epsilon) \right) / \vec{v}^T\vec{u}_0(\epsilon)}$. Upon dividing \cref{eq:h_epsilon} by $\epsilon$, taking the limit $\epsilon \rightarrow 0$ and $N \rightarrow \infty$ where $ \epsilon = 1/N$, and introducing \review{$h(x,t) = \text{lim}_{\epsilon \rightarrow 0} \left\lbrace h_{\epsilon}(x,t) \right\rbrace$, $\hat{V} = \text{lim}_{\epsilon \rightarrow 0} \left\lbrace V(\epsilon) / \epsilon \right\rbrace $, and $\hat{D} = \text{lim}_{\epsilon \rightarrow 0} \left\lbrace D(\epsilon) \right\rbrace $} we have the following linear advection-diffusion PDE
\begin{equation}\label{macroPDE}
\dfrac{\partial h}{\partial t} + \hat{V} \dfrac{\partial h}{\partial x} = \hat{D}  \dfrac{\partial^2 h}{\partial x^2},
\end{equation}
for $x \in [0,\infty)$, with boundary conditions $\hat{D} h_x(0,t) + \hat{V} h(0,t) = \lim_{x \rightarrow \infty} \{\hat{D} h_x(x,t) + \hat{V} h(x,t) \} = 0$ and initial conditions $h(x,0) = \delta(x)$. \Cref{macroPDE} is a macroscopic PDE for the evolution of the displacement (defined as the winding distance) of a particle on a network $\mathcal{G}$, valid in the long-time limit. \review{Note that the scalar coefficient of $\vec{u}_1(\epsilon)$ in \cref{eq:expansion} is $\epsilon f_x(x,\hat{t})$. Upon rescaling time $t = \epsilon \hat{t}$, it is clear from \cref{eq:f} that $f_x(x,t)$ is independent of $\epsilon$, and we find that $\lim_{\epsilon \rightarrow 0}\left\lbrace \epsilon f_x(x,t)\right\rbrace = 0$. The entries of $\vec{u}_1(0)$ given in \cref{eq:u1} are finite. Therefore, as $\epsilon \rightarrow 0$, there is no contribution to $\vec{P}(x,t)$ from the term $\epsilon f_x(x,t)\vec{u}_1 (\epsilon)$.} The terms $\hat{V} h_x$ and $\hat{D} h_{xx}$  in \cref{macroPDE} represent the effective drift and diffusion contributions, respectively, to the distribution of the displacement of a particle. The transport coefficients $\hat{V}$ and $\hat{D}$ determine the strengths of these contributions as a function of the networked topology. We use the expressions for $\mathbf{A}$, $\mathbf{B}$ and $\mathbf{C}$ in  \cref{eq:matrices}, \review{as well as $\vec{u}_0(\epsilon)$ and $\vec{u}_1(\epsilon)$ in \cref{eq:u0} and \cref{eq:u1}, respectively,} to calculate the effective transport coefficients as
\begin{subequations} \label{coefficients}
\begin{eqnarray}
\hat{V} &=&  \rho^+ - \rho^- , \\
\hat{D}&=& ~\dfrac{1}{2} - \dfrac{1}{6} \left( \rho^+ - \rho^- \right)^2.
\end{eqnarray}
\end{subequations}
From \cref{macroPDE} we calculate expressions for the first two moments of the displacement as
\begin{subequations}\label{longtime}
\begin{eqnarray}
\langle x(t) \rangle &\sim& \hat{V} t = \left( 1 - \dfrac{2}{\bar{d}} \right) t, \label{longtime_a}\\
\langle x^2(t) \rangle &\sim& \hat{V}^2 t^2 + 2\hat{D}t = \left(1 - \dfrac{2}{\bar{d}}\right)^2 t^2 + \left[ 1 - \dfrac{1}{3}\left( 1 - \dfrac{2}{\bar{d}} \right)^2 \right] t, \label{longtime_b}
\end{eqnarray}
\end{subequations}
where $\bar{d}$ is the average degree of the network. The ballistic term $\hat{V}^2 t^2$ in \eqref{longtime_b} is nonzero when $\bar{d} > 2$, that is, the average degree of $\mathcal{G}_L$ must be strictly greater than two for the network to induce ballistic behaviour in the winding distance of a diffusive particle. 

\subsection{Full distribution of displacement}

\begin{figure}[tb]\label{fig:fulldist}
\centering
$\begin{array}{c}
\includegraphics[scale = 0.465]{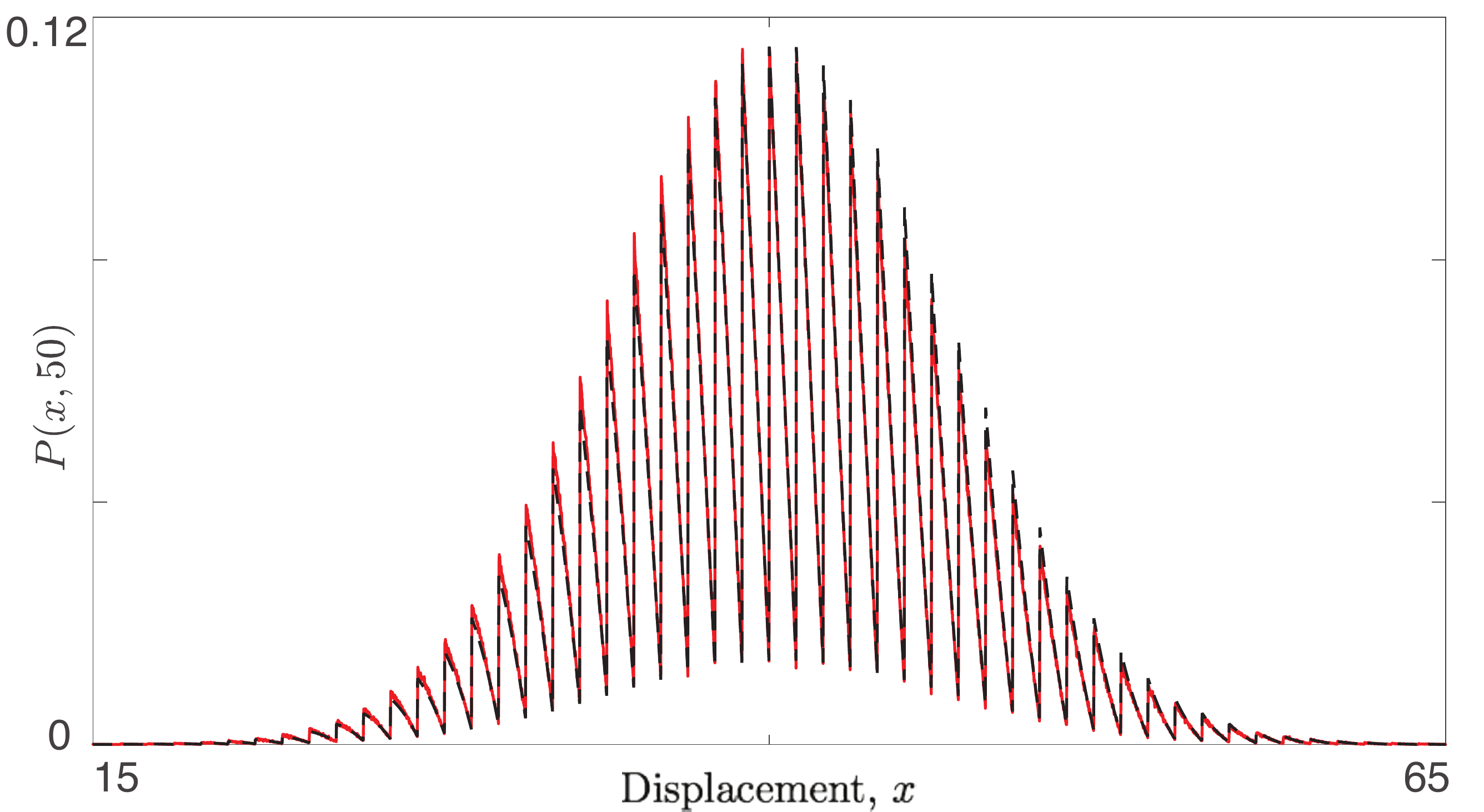}
\end{array}$
\caption{The displacement of a particle on $\mathcal{G}_{10}$ estimated using $10^6$ realisation of a Brownian particle is shown by the solid red curve. The analytical expression for the displacement of a particle on the network $\mathcal{G}_{10}$ in the long-time limit is shown by the dashed black curve. All realisations of a Brownian motion were numerically integrated using the Euler-Maruyama method with timestep $\Delta t = 10^{-5}$, $D = 1/2$ and $L=1$.  }
\end{figure}

We are not in fact limited to approximating the first two moments of the displacement but can also obtain an analytical approximation to the full distribution. Recall that the distribution of the displacement of a particle whose position evolves according to the CTMC is equivalent to the same distribution as that of a Brownian particle in the limit $\epsilon \rightarrow 0$ and $N \rightarrow \infty$ such that $\epsilon = 1/N$. Note that \review{$\vec{P}(x,t) \approx h_{\epsilon}(x,t) \vec{u}_0(\epsilon)$, where $h(x,t) = \lim_{\epsilon \rightarrow 0} \left\lbrace h_{\epsilon}(x,t) \right\rbrace$ satisfies} \cref{macroPDE} which has solution
\begin{equation}\label{eq:h}
h(x,t) = \dfrac{1}{\sqrt{\pi \hat{D} t}} \text{exp}\left({-\dfrac{(x-\hat{V}~t)^2}{4\hat{D}t}}\right) + \dfrac{\hat{V}}{2\hat{D}} \text{exp}\left({-\dfrac{\hat{V}x}{\hat{D}}}\right) \text{erfc} \left( \dfrac{x+\hat{V}t}{2\sqrt{\hat{D}t}}\right),
\end{equation}
where $\text{erfc}(\cdot)$ is the complimentary error function \cite{Monthus_BetheLattice}, \review{and $\vec{u}_0(\epsilon)$ is given in \cref{eq:u0}. We introduce the function $u(x) = \lim_{\epsilon \rightarrow 0} \{ \epsilon^{-1} \left[ \vec{u}_0(\epsilon) \right]_n \}$, where $x=\epsilon n$} and find that
\begin{equation}\label{eq:u}
u(x) = 2\rho^+ (1-x) + 2\rho^- x,
\end{equation}
for $x \in [0,1)$ and $u(1) = 1$. Extending the function $u(x)$ to be periodic on the half line with unit period, $u(x+1) = u(x)$ for all $x \in (0,\infty)$, provides us with a function that accounts for corrections on the microscopic scale to the displacement predicted from the macroscopic equation for $h(x,t)$. Then the full distribution is approximated by $P(x,t) \approx  h(x,t)u(x)$. For $\mathcal{G}_{10}$, we simulate $10^6$ realisations of a Brownian particle terminating at time $t=50$. We use these realisations to empirically plot the distribution of displacement (\cref{fig:fulldist}; red solid curve) and compare with the analytical approximation using the solution $h(x,t)$ in \cref{eq:h} and the periodically extended function $u(x)$ in \cref{eq:u} (\cref{fig:fulldist}; black dashed curve). The two curves match well which suggests that the temination time $t=50$ is sufficiently large for the transport process to have reached its equilibrium behaviour for the network $\mathcal{G}_{10}$. However, in general how does the topology of the network affect the time taken to reach equilibrium?

\subsection{Timescale for transition to equilibrium}

\begin{figure}[tb]
\centering
$\begin{array}{c}
\includegraphics[scale=1.05]{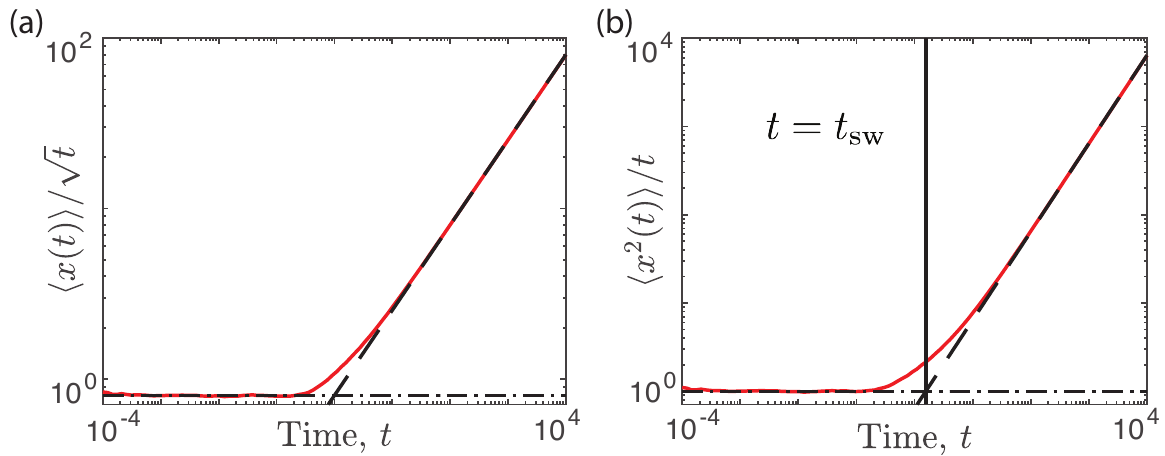}
\end{array}$
\caption{(a): The first moment of displacement of a particle on the network $\mathcal{G}_{10}$ estimated from $10^4$ realisations of a Brownian particle (solid red curve). Analytical results of the first moment in the short-time (dot-dashed line) and long-time (dashed line) limits. (b): The MSD of a particle on the network $\mathcal{G}_{10}$ estimated from $10^4$ realisations of a Brownian particle (solid red curve). Analytical results of the MSD in the short-time (dot-dashed line) and long-time (dashed line) limits. The predictive timescale $t_{\text{sw}}$ is identified with the vertical solid line. All simulations of a Brownian motion were numerically integrated using the Euler-Maruyama method with timestep $\Delta t = 10^{-5}$, $D=1/2$ and $L=1$.  }\label{fig:2moments}
\end{figure}

From \cref{longtime} we already have the dominant contributions to the first two moments of displacement for long times. Now consider short times, $t$, sufficiently short such that the particle has a displacement less than one, or equivalently has not reached the first branching point on the displacement tree. On this timescale, the particle undergoes unbiased Brownian motion on the half line, for which the first two moments of displacement are
\begin{equation}\label{shorttime}
\langle x(t) \rangle \sim  \dfrac{2 \sqrt{t}}{\sqrt{\pi}}, \hspace{10 mm} \langle x^2(t) \rangle \sim  t.
\end{equation}
The dominant terms of the long-time and short-time analytical asymptotics in \cref{longtime} and \cref{shorttime}, respectively, are plotted in \Cref{fig:2moments} (dashed and dot-dashed lines). Equating the dominant contributions to the MSD on both the long and short timescales, $\hat{V}^2 t^2$ and $t$, respectively, we obtain a prediction of the timescale, $t_{\text{sw}}$, upon which the displacement of a particle transitions from characteristically diffusive to ballistic:
\begin{equation}
t_{\text{sw}} \sim  \left( \rho^+ - \rho^-\right)^{-2} =   \left( 1 - \dfrac{2}{\bar{d}}\right)^{-2}, \label{eq:tswitch}
\end{equation}
where $\bar{d} = 2|\mathcal{E}_L|/|\mathcal{V}_L|$ is the average degree of the network. The timescale, $t_{\text{sw}}$, is given by the intersection of the dashed and dot-dashed lines in \Cref{fig:2moments}(b) (vertical line). 

\begin{figure}[tb]\label{fig:100}
\centering
$\begin{array}{c}
\includegraphics[scale=0.3]{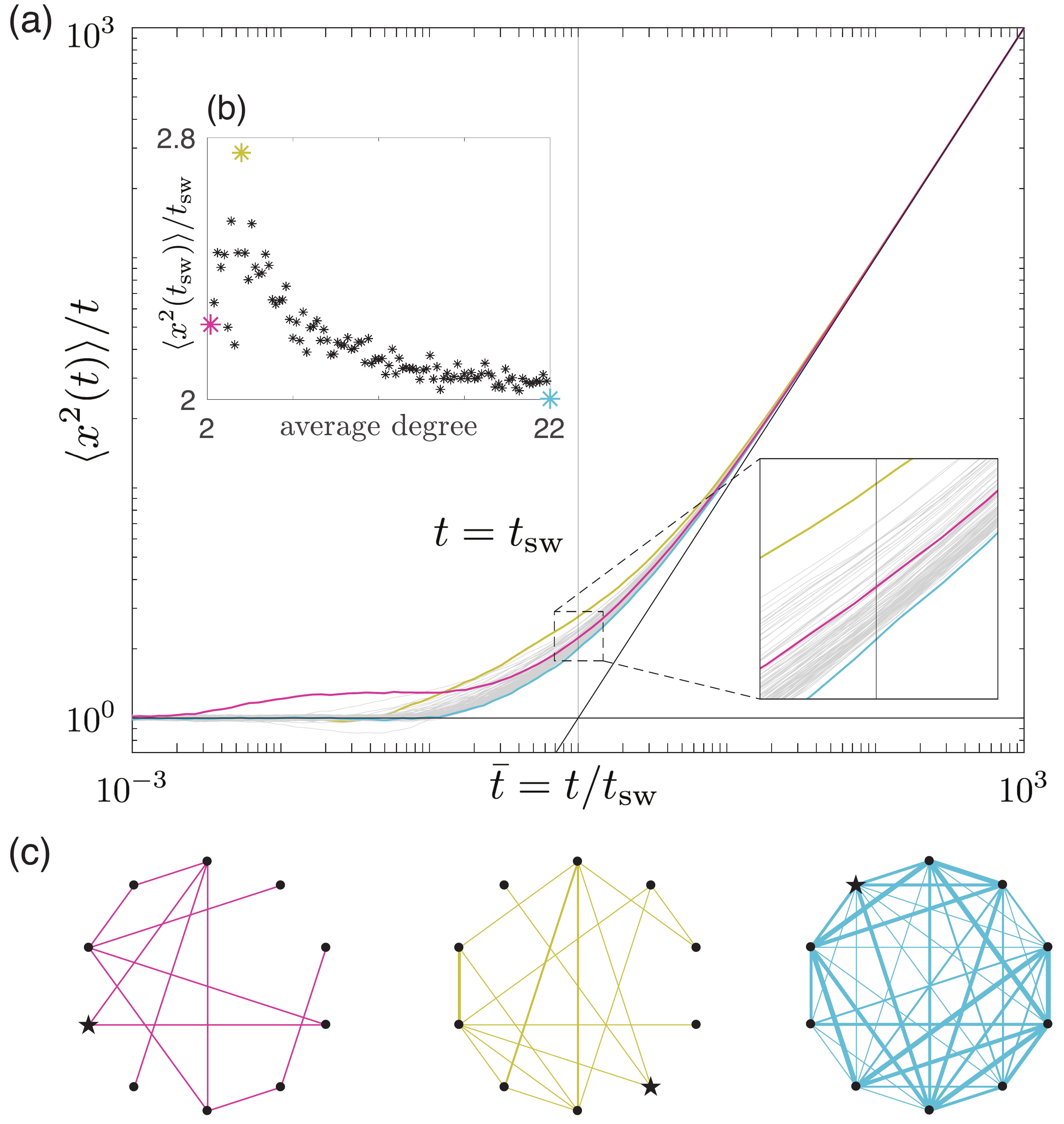}
\end{array}$
\caption{(a): The MSD of a particle on $100$ randomly generated networks, estimated using $10^4$ realisations of a Brownian motion on each network. The inset shows the MSD curves for all $100$ networks around the predictive timescale $t = t_{\text{sw}}$, which is identified by the vertical line. The short-time and long-time analytical expressions for the MSD are shown in solid black lines. (b): The $y$ intercepts of the inset in (a) are plotted against the average degree of each network. (c): Three networks of note are shown where the width of the edges corresponds to the number of multiple edges between the two adjacent vertices. The initial vertex in each network is represented as a star. All simulations were calculated through numerical integration of a Brownian motion using the Euler-Maruyama method with timestep $\Delta t = 10^{-5}$, $D = 1/2$ and $L=1$. }
\end{figure}

To investigate the predictive capacity of the timescale $t_{\text{sw}}$, we consider the MSD over an ensemble of randomly generated aperiodic networks. The MSD of a particle on a general network has dominant contribution $\langle x^2(t) \rangle \sim \hat{V}^2 t^2$ for long times. The rescaling $\bar{t} = t/t_{\text{sw}}$ allows for the MSD of a particle over an ensemble of topologies to be compared on the same temporal axis, as $\langle x^2(t) \rangle / t \sim \bar{t}$. In \Cref{fig:100}(a) the MSD of a particle over an ensemble of $100$ networks is presented. The $100$ networks each have $10$ vertices and a range of $11$ to $110$ edges\footnote{The sampling procedure is as follows; select a desired number of edges and vertices; draw two distinct vertices uniformly at random and join them via an edge; repeat the previous step until the required number of edges is reached; check if the network is periodic or disconnected, if so reject and start over, if not take the network as a realisation. This procedure allows for multiple edges but avoids self loops.}. For each network, the particle starts at the same vertex for all realisations, but that fixed vertex is randomly selected and highlighted in \Cref{fig:100}(c) with a star\footnote{The initial vertex is fixed among realisations of the transport process, so that the displacement tree has a fixed root and the effects of the initial conditions on transition times are not averaged out.}.

On visual inspection of the data in \Cref{fig:100} we see that the grey curves rarely intersect as they collapse onto the curve $\langle x^2(t) \rangle / t = \bar{t}$. As such, the curves that intersect the line $t = t_{\text{sw}}$ (see \Cref{fig:100}(a) inset) with a lower $y$ intercept will transition to ballistic behaviour before those curves with a greater $y$ intercept. Thus, we plot the $y$ intercept, $\langle x^2 (t_{\text{sw}}) \rangle / t_{\text{sw}}$, for all $100$ networks against the average degree of the network in \Cref{fig:100}(b). We see a sharper correlation between average degree and the $y$ intercept for networks with a higher average degree. This suggests that $t_{\text{sw}}$ is a good prediction of the crossover time to ballistic motion when the average degree is large (\Cref{fig:100}: the network highlighted in blue). However, for networks with a lower average degree the predictive capacity of $t_{\text{sw}}$ is less clear; we highlight networks in magenta\footnote{The magenta network has an average degree $\bar{d} = 2.2$ however the fixed initial condition is the vertex of degree two highlighted with a star in \Cref{fig:100}(c). The environment local to the initial vertex has a significantly higher average degree than the entire network, which results in a curve that deviates from the short-time behaviour earlier (see the magenta curve in \Cref{fig:100}(a)). The dip in the gradient of the magenta curve occurs later, when the particle has explored the global environment, which has a lower average degree than the local environment.} and yellow in \Cref{fig:100} that both have low average degrees yet the values of $\langle x^2 (t_{\text{sw}}) \rangle / t_{\text{sw}}$ are very different (\Cref{fig:100}(b)). For topologies with a low average degree the timescale to transition from diffusive to ballistic behaviour is highly sensitive to the particular network and, as such, $t_{\text{sw}}$ may be less reliable as a predictor of the crossover to ballistic motion.

\section{\label{FracPDES}Anomalous diffusion on networks}

Thus far, the only transport process we have considered is Brownian motion. An alternative modelling framework for particle transport is a position jump process, where the position of a particle is updated instantaneously after random intervals in time. A position jump process can be formally described as a continuous time random walk (CTRW) where the jump distance is sampled from $\lambda (x)$, the jump distribution, and the time between jump events is sampled from $\omega (t)$, the waiting time distribution. For a review of transport properties of these CTRWs see \cite{Metzler2000}. The freedom to prescribe any jump and waiting time distributions allows a CTRW to describe a broad range of physical processes. Consider a waiting time distribution that for long times has asymptotic power law behaviour, $\omega (t) \sim  t ^{-\left( 1+\alpha \right)}$ for some $\alpha$. If $\lambda (x)$ has zero mean and finite variance then for $\alpha \geq 1$, the position of a particle evolving according to the CTRW on the real line in the long-time limit is equivalent to the position of a Brownian particle and hence has a MSD of $\langle x^2(t) \rangle \propto t$. However, for $0 < \alpha < 1$, $\omega (t)$ is heavy tailed and the mean waiting time between jump events diverges. Such transport processes for large times have a MSD on the real line of $\langle x^2(t) \rangle \propto t^{\alpha}$, and are known as anomalously diffusive. 

Using the displacement tree (\Cref{sec:displacement_tree}), the winding distance of anomalously diffusive transport on networks can be identified with displacement of the same transport process on the half line subject to periodic bias. As before, we consider a multiple scales approach and discretise the half line into intervals, $[k,k+1]$ for integers $k \geq 0$, each of $N+1$ lattice sites separated by a distance $\epsilon = 1/N$. The position of a particle on the lattice evolves according to a CTRW rather than a CTMC because the waiting times for jumps between adjacent sites are no longer exponentially distributed. Similar to \Cref{PDEs} we now select the fractional transition rates\footnote{The mean time for a jump to occur on the lattice diverges, as such we do not consider transition rates as we would for a CTMC. Instead we introduce a fractional transition rate \cite{HilferPRE1995}.} to represent the periodic bias felt at the ends of the intervals $[k,k+1]$. Thus, for a particle to exit a lattice site with the index $kN$ to the right with probability $\rho^+$ we select the transition rates $\mu_{kN\rightarrow kN+1} = \rho^+ /  \epsilon^2 $ and $\mu_{kN\rightarrow kN-1} = \rho^- / \epsilon^2 $, and all other lattice sites $n$ have symmetric transition rates $\mu_{n \rightarrow n+1} = \mu_{n \rightarrow n-1} = 1/ \left( 2 \epsilon^2 \right)$.

Let $p_n(t)$ be the probability a particle occupies lattice site $n$ at time $t$. The temporal evolution of the probability $p_n(t)$ evolves according to the fractional master equation \cite{HilferPRE1995}, a generalisation of the master equation seen in \Cref{PDEs}. We define the operator $\partial ^{\alpha} / \partial t^{\alpha}$ as the Riemann-Liouville fractional derivative
\begin{equation}
\dfrac{\partial^{\alpha}}{\partial t^{\alpha}} \left\lbrace y(t) \right\rbrace = \dfrac{1}{\Gamma \left( 1-\alpha \right)} \dfrac{\partial}{\partial t} \int_{0}^t \dfrac{y(t')}{\left( t-t'\right)^{\alpha}} \mathrm{d}t'.
\end{equation}
The fractional master equation for the probabilities $p_n(t)$ is 
\begin{equation}\label{eq:FME}
\dfrac{\partial^{\alpha} p_n}{\partial t^{\alpha}} = \mu_{(n-1)\rightarrow n} p_{n-1} + \mu_{(n+1)\rightarrow n} p_{n+1} - \left( \mu_{n \rightarrow (n-1)} + \mu_{n \rightarrow (n+1)} \right) p_n,
\end{equation}
where $n \geq 0$. As before, the particle is initially at the origin, therefore $p_0(0)=1$ and $p_i(0)=0$ for all $i \geq 1$. To account for the reflective boundary condition at the origin, we set $\mu_{-1 \rightarrow 0} = \mu_{0 \rightarrow -1} = 0$ and $\mu_{0 \rightarrow 1} = 1/  \epsilon^2 $. Introducing the spatial coordinate $x = \epsilon N$ where $\epsilon \ll 1$, we write $p_n(t) = P_n(x,t)$ so that the probability density for the position of a particle at time $t$ depends on both the fast scale $n$ and the slow scale $x$. Treating the two variables $x$ and $n$ independently and rewriting the fractional master equation \cref{eq:FME} in terms of $P_n(x,t)$ gives the finite set of equations
\begin{align}\label{eq:FME_periodic}
\dfrac{\partial^{\alpha} P_i}{\partial t^{\alpha}}(x,t) =& ~\mu_{\overline{i-1}\rightarrow i} P_{\overline{i-1}}(x-\epsilon,t) + \mu_{\overline{i+1}\rightarrow i} P_{\overline{i+1}}(x+\epsilon,t) \\
& - \left( \mu_{i \rightarrow \overline{i+1}} + \mu_{i \rightarrow \overline{i-1}}\right) P_i(x,t), \notag
\end{align}
for $i \in \{1 , \ldots , N \}$ and $x \in [0, \infty)$, where $\overline{i \pm 1} = \left( {i \pm 1} - 1 \text{ mod } N \right) + 1$. For the periodic interval, as before the transition rates out of all sites are symmetric and equal to $1/\left(2 \epsilon^2\right)$, other than the rates $\mu_{N \rightarrow 1} = \rho^+ / \left( \epsilon^2\right)$ and $\mu_{N \rightarrow N-1} = \rho^- / \left( \epsilon^2\right)$. Introducing $\vec{P}(x,t) = \left[ P_1(x,t) , \ldots , P_N(x,t) \right]^T$ and expanding the right-hand side of \cref{eq:FME_periodic} about $x$, we find
\begin{equation}\label{FracPDE}
\epsilon^2 \dfrac{\partial^{\alpha} \vec{P}}{\partial t^{\alpha}}(x,t) = \mathbf{A} \vec{P} + \epsilon \mathbf{B} \dfrac{\partial \vec{P}}{\partial x} + \epsilon^2 \mathbf{C} \dfrac{\partial^2 \vec{P}}{\partial x^2} + \mathcal{O} \left( \epsilon^3 \right),
\end{equation}
where the matrices are as given in \cref{eq:matrices}. Let $h(x,t)$ denote the macroscopic probability density function for a particle to have a winding distance of $x$ at time $t$. The analysis proceeds identically to \Cref{PDEs} and we derive the following fractional advection-diffusion PDE
\begin{equation}
\dfrac{\partial^{\alpha} h}{\partial t^{\alpha}}(x,t) + \hat{V} \dfrac{\partial h}{\partial x} = \hat{D}  \dfrac{\partial^2 h}{\partial x^2},
\end{equation}
for $x \in [0,\infty)$, with boundary conditions $\hat{D} h_x(0,t) + \hat{V} h(0,t) = \lim_{x \rightarrow \infty} \{\hat{D} h_x(x,t) + \hat{V} h(x,t) \} = 0$ and initial conditions $h(x,0) = \delta(x)$. The transport coefficients $\hat{V}$ and $\hat{D}$ are given in \cref{coefficients}. This macroscopic fractional PDE is valid for large times and the first two moments of displacement are calculated to be 
\begin{subequations}\label{SubDiff}
\begin{align}
\langle x(t) \rangle &\sim \dfrac{\hat{V}t^{\alpha}}{\Gamma \left( 1+\alpha \right)} = \dfrac{1}{\Gamma(1+\alpha)} \left( 1 - \dfrac{2}{\bar{d}}\right) t^{\alpha}, \\
\langle x^2(t) \rangle &\sim  \dfrac{2\hat{V}^2 t^{2\alpha}}{\Gamma\left( 1 + 2\alpha \right)} + \dfrac{2 \hat{D} t^{\alpha}}{\Gamma \left( 1 + \alpha \right)} \label{SubDiff_MSD}\\
&= \dfrac{2}{\Gamma \left(1+2\alpha \right)} \left( 1 - \dfrac{2}{\bar{d}}\right)^2 t^{2 \alpha} + \left[ 1 - \dfrac{1}{3} \left( 1 - \dfrac{2}{\bar{d}}\right)^2 \right] \dfrac{t^{\alpha}}{\Gamma \left( 1 + \alpha \right)}. \notag
\end{align}
\end{subequations}

\Cref{SubDiff} demonstrates that anomalous diffusion on a network can induce a variety of transitional behaviours. Consider the analytical expression in the long-time limit for the MSD in \cref{SubDiff_MSD}. At large times the dominant contribution is proportional to $t^{2 \alpha}$ when $\bar{d} > 2$. If $0<\alpha<1/2$, the winding distance of the particle in the long-time limit remains sub-diffusive but with an increased exponent of $2\alpha$. If $\alpha = 1/2$, the winding distance becomes classically diffusive. For $1/2 < \alpha < 1$ the winding distance transitions to sub-ballistic or super-diffusive. Generalising to anomalously diffusive transport processes demonstrates that in order to capture the effect of network topology on the winding distance of a particle one must also incorporate details of the transport process itself.

As in \Cref{PDEs} we shall compare the first two moments of displacement on both long and short timescales. On short timescales, where the particle has not left the first edge, the displacement of a particle is equivalent to the displacement of a particle whose position is described by a fractional diffusion equation on the half line. The first two moments of displacement \cite{MetzlerKlafter1999} on short timescales are 
\begin{equation}
\langle x(t) \rangle \sim \dfrac{t^{\alpha/2}}{\Gamma \left( 1 + \alpha/2 \right)}, \hspace{10mm} \langle x^2(t) \rangle \sim \dfrac{ t^{\alpha}}{\Gamma \left( 1+ \alpha \right)}.
\end{equation} 
We compare the dominant short-time and long-time behaviours for the MSD to estimate the transition timescale
\begin{equation}\label{eq:fracswitch}
t_{\text{sw}} =  \left[ \dfrac{\Gamma\left( 1 + 2 \alpha\right)}{2 \Gamma\left( 1+ \alpha\right)} \left( 1 - \dfrac{2}{\bar{d}}\right)^{-2} \right]^{1/\alpha},
\end{equation}
where for $\alpha=1$ we recover \cref{eq:tswitch}.

\begin{figure}[tb]\label{fig:FracAlpha}
\centering
$\begin{array}{c}
\includegraphics[scale=0.57]{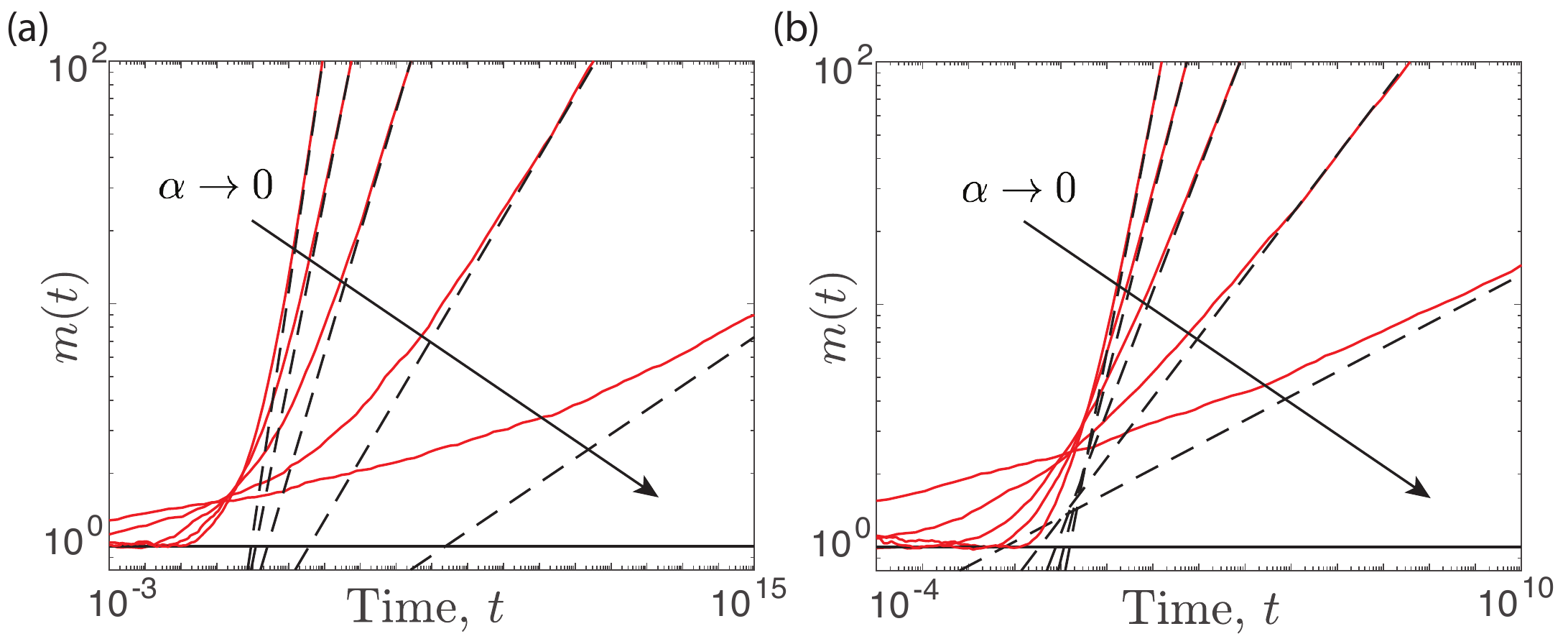}
\end{array}$
\caption{(a): The rescaled MSD of a particle on the network $\mathcal{G}_{3}$ estimated from $10^4$ realisations of the CTRW (solid red curves) for an ensemble of $\alpha$ values. Analytical results of the rescaled MSD in the short-time (solid black line) and long-time (dashed black lines) limits. (b): The rescaled MSD of a particle on the network $\mathcal{G}_{10}$ estimated from $10^4$ realisations of the CTRW (solid red curves) for an ensemble of $\alpha$ values. Analytical results of the rescaled MSD in the short-time (solid black line) and long-time (dashed black lines) limits. The selected values of $\alpha$ are $\{0.1,0.25,0.50,0.75,1.0 \}$. For $\alpha=1.0$ the MSD was calculated using the Brownian motion simulations from \Cref{Fig2.1} and \Cref{fig:2moments}. For the other values of $\alpha$, realisations of the CTRW were simulated with a normally distributed jump length distribution with zero mean and variance $\sigma^2$, and a Pareto waiting time distribution with $t_m = \left( \sigma^2/\Gamma \left( 1 - \alpha \right) \right)^{1/\alpha}$. For the values $\alpha = \{0.1,0.25,0.50,0.75\}$ the standard deviations used are $\sigma = \{10^{-2},10^{-2},10^{-2},10^{-2.75} \}$, respectively. The edges are all of unit length.}
\end{figure}

Consider the limit where $\alpha \rightarrow 0$, which corresponds to a transport process where the waiting time distribution between jump events becomes increasingly heavy tailed. Expanding \cref{eq:fracswitch} for small $\alpha$ yields
\begin{equation}\label{eq:fracswitch_expand}
t_{\text{sw}} \sim \left( 1 + \dfrac{\pi^2}{4} \alpha \right) \exp \left[ - \left( \gamma + \dfrac{\log(2)}{\alpha} + \dfrac{2 \log\left( 1-2/\bar{d}\right)}{\alpha} \right) \right],
\end{equation}
where $\gamma$ is the Euler-Mascheroni constant. In the limit $\alpha \rightarrow 0$ there are two cases for the limiting behaviour of the timescale $t_{\text{sw}}$. If the average degree of the network satisfies $\bar{d} < 2 \left( 2 + \sqrt{2} \right)$ then $t_{\text{sw}} \rightarrow \infty$, and if $\bar{d} \geq 2 \left( 2 + \sqrt{2} \right)$ then $t_{\text{sw}} \rightarrow 0$. To explain this, in \Cref{fig:FracAlpha} we estimate the rescaled MSD, $m(t) = \Gamma(1+\alpha) \langle x^2(t) \rangle / t^{\alpha}$, from simulations of the CTRW on the networks $\mathcal{G}_{3}$ and $\mathcal{G}_{10}$ over an ensemble of values for $\alpha$. The waiting times are Pareto distributed, $w(t) = \alpha t_m^{\alpha} / t^{\alpha+1}$ \review{for $t \geq t_m$}, and the jump lengths are normally distributed with zero mean and variance $\sigma^2$. \review{We choose the minimum value of the waiting time, $t_m = \left( \sigma^2 / \Gamma \left(1 - \alpha \right) \right)^{1/\alpha}$, to ensure the simulations of the CTRW agree with the multiscale analysis.}\footnote{\review{The Laplace transform of $w(t)$ for small values of $s$ is given by $\tilde{w}(s) \approx 1 - \Gamma(1-\alpha) t_m^{\alpha} s^{\alpha}$, thus the cofficient of $s^{\alpha}$ provides a timescale, $\tau$, upon which the Pareto distribution decays, where $\tau^{\alpha} = \Gamma(1-\alpha) t_m^{\alpha}$. Following the derivation of the fractional diffusion equation seen in \cite{Metzler2000} the fractional diffusion coefficient is equal to $\sigma^2 / \left( 2 \tau^{\alpha} \right)$. The hopping rates in \cref{eq:FME} are selected such that as $\epsilon \rightarrow 0$, the resulting fractional diffusion coefficient is equal to $1/2$. Thus, we select $\tau^{\alpha} = \sigma^2$, or equivalently $t_m = \left( \sigma^2 / \Gamma \left(1 - \alpha \right) \right)^{1/\alpha}$.}} For $\mathcal{G}_3$ the average degree is less than $2(2+\sqrt{2})$ and so $t_{\text{sw}} \rightarrow \infty$ as $\alpha \rightarrow 0$ (see intersection points with $m(t)=1$ in \Cref{fig:FracAlpha}(a)). On inspection we see that the time to converge to equilibrium behaviour increases as expected when $\alpha \rightarrow 0$. In contrast, the average degree of $\mathcal{G}_{10}$ is greater than $2(2+\sqrt{2})$, meaning $t_{\text{sw}} \rightarrow 0$ as $\alpha \rightarrow 0$ (see intersection points with $m(t)=1$ in \Cref{fig:FracAlpha}(b)). But if we inspect the curve in \Cref{fig:FracAlpha}(b) corresponding to $\alpha = 0.1$, we see it has transitioned away from the short-time behaviour significantly for times as small as $t=10^{-4}$ and is only just converging to equilibrium behaviour at $t=10^{10}$, a difference of $14$ orders of magnitude. Thus, for small values of $\alpha$, it is difficult to consider a single value (such as $t_{\text{sw}}$) that represents the time taken to transition to equilibrium behaviour as the change occurs gradually over a large window of time. That said, for values of $\alpha$ closer to one the transition occurs more rapidly, and $t_{\text{sw}}$ provides a good predictive timescale.


\section{Summary and discussion}\label{sec:summary}

In this work, we have investigated the influence of network topology on the evolution of the displacement of a particle whose position evolves according to either a diffusive or an anomalously diffusive transport process. In \Cref{sec:displacement_tree} we define our notion of displacement of a particle as the winding distance, which is a measure of how far a particle has travelled as it winds around the network. To investigate the evolution of the winding distance we introduced a topological structure we term the displacement tree and noted that the winding distance of a particle is equivalent to the displacement of a particle along the tree. Furthermore, in the long-time limit, a link between the displacement of a particle on the tree and the displacement of a particle on the half line with periodic bias was highlighted. The simpler transport process on the half line has a periodic bias with strength $\rho^+_{\infty}$, a quantity that can be calculated in terms of the average degree of the network.

In \Cref{PDEs} a multiple scales approach from \cite{ChapmanShabala17} was used to derive an advection-diffusion PDE for the displacement of a particle whose position evolves according to a Brownian motion on the half line with periodic bias. The advective and diffusive transport coefficients are calculated in terms of the periodic bias $\rho^+_{\infty}$. From these coefficients a topological condition was derived for whether the long-time behaviour of the winding distance switches to ballistic motion or remains diffusive. By comparing the MSD for the short-time and long-time limits, a prediction of the timescale upon which a network induces the qualitative switch from diffusive to ballistic behaviour was obtained. For an ensemble of $100$ randomly generated networks we found that the predictive timescale performs better for networks with a higher average degree. 

Finally, in \Cref{FracPDES}, we discussed an extension to a class of anomalously diffusive transport processes using the CTRW framework. Through the use of the fractional master equation we derived a time fractional advection-diffusion PDE for the evolution of the displacement of a particle on the half line with periodic bias. Our results showed that the characteristic nature of the long-time behaviour depends upon the transport process itself. We found that, depending on the value of the exponent $\alpha$, the displacement of a particle in the equilibrium limit can remain sub-diffusive, become diffusive, or become super-diffusive. Similarly to \Cref{PDEs} we derived a predictive timescale upon which these transitions occur and explored its validity as a function of the parameter $\alpha$, the exponent in the heavy-tailed waiting time distribution of the CTRW.

This work introduced the winding distance, a new measure of displacement for particles on networked topologies. Using the winding distance to investigate both diffusive and anomalously diffusive transport processes has highlighted that both the topology of the network and the nature of the transport process itself play critical roles in the evolution of particle displacement. Active transport through networked environments is a critical feature of many biological systems, such as the cytoplasmic streaming seen within \textit{Drosophilia} oocytes \cite{Ganguly_PNAS12,Nieuwberg_eLife17}, the transport of bronchial mucus \cite{Vasquez16}, and the transport of a broad range of biomolecules along microtubular networks \cite{Franker13,Welte04}. Explicitly incorporating network topology within these biological transport systems requires the theoretical study of models for stochastic transport in networked environments \cite{Kahana08, Bresloff_RMP}. Therefore, future work will extend our investigation of the winding distance to transport processes with a directional bias along each of the edges in a network.


\appendix
\section{Simulating the transport processes} \label{app:simulations} Here we present the algorithm used to compute realisations of a Brownian particle on a network $\mathcal{G}=\{ \mathcal{V}, \mathcal{E} \}$, using the Euler-Maruyama method.
\begin{algorithmic}[1]

\State{Assign orientations to each edge in the network, select an initial vertex for the particle, set $t=0$, the termination time $T$ and the time step $\Delta t$.}

  \State{Sample an adjacent edge, $e$, to the initial vertex uniformly at random. Let $X(0) = 0$ or $X(0)=L_e$ depending on the orientation of edge $e$.}
  
\While{$t <  T$}

\If{$t = 0$} update the position as follows:
  \begin{equation}\label{A1}
  X\left( t+\Delta t \right) = X(t) + \sqrt{2D \Delta t}~ |\xi|,
  \end{equation}
  where $\xi \sim \mathcal{N}(0,1)$.
\Else{
~update the position as follows:
  \begin{equation}\label{A2}
  X\left( t+\Delta t \right) = X(t) + \sqrt{2D \Delta t}~ \xi,
  \end{equation}
  where $\xi \sim \mathcal{N}(0,1)$.}
\EndIf

\If{ $X(t+\Delta t) < 0$} sample a new edge $e^*$, uniformly at random from the adjacent edges to edge $e$ (not including $e$ itself) at the end with position $0$.
\If{ the particle enters the new edge $e^*$ at the end with position $0$} update the new position as
\begin{equation}
X(t+\Delta t ) \leftarrow - X(t+\Delta t),
\end{equation}
\ElsIf{ the particle enters the new edge $e^*$ at the end with position $L_{e^*}$} update the new position as
\begin{equation}
X(t+\Delta t ) \leftarrow L_{e^*} + X(t+\Delta t),
\end{equation}
\EndIf 

\State{Update $e=e^*$.}

\ElsIf{$X(t+\Delta t) > L_e$} sample a new edge $e^*$, uniformly at random from the adjacent edges to edge $e$ (not including $e$ itself) at the end with position $L_e$.
\If{ the particle enters the new edge $e^*$ at the end with position $0$} update the new position as
\begin{equation}
X(t+\Delta t ) \leftarrow X(t+\Delta t) - L_e,
\end{equation}
\ElsIf{ the particle enters the new edge $e^*$ at the end with position $L_{e^*}$} update the new position as
\begin{equation}
X(t+\Delta t ) \leftarrow L_{e^*} + L_e - X(t+\Delta t), 
\end{equation}
\EndIf 
\State{Update $e=e^*$.}
\EndIf
\State{Update the time $t = t + \Delta t$.}
\EndWhile

\end{algorithmic}

For the CTRWs seen in \Cref{FracPDES} we select the jump distribution to be normal with zero mean and standard deviation $\sigma$. The above algorithm is easily adapted to simulate the CTRW on a network. \review{Instead of fixed time steps, $\Delta t$, update the time $t$ with random waiting times from the Pareto distribution, and replace $\sqrt{2D \Delta t}$ in equations \cref{A1} and \cref{A2} with $\sigma$. The CTRW simulations must agree with the multiscale analysis in \Cref{FracPDES}, on both long and short timescales. In particular, on timescales sufficiently short that a particle has not left the initial edge. The multiscale analysis discretises the unit interval into $N+1$ lattice sites. As $N \rightarrow \infty$ the width of each site, $\epsilon = 1/N$, tends to zero, and a particle will undergo a large number of jumps before traversing an entire interval. Therefore, for the CTRW to agree with the multiscale analysis on short timescales, we take $\sigma^2 \ll 1$, such that many jumps must occur before a particle traverses an edge.}

\section{Periodic networks} \label{app:periodic}

Recall the sequence of biased probabilities $\{ \rho^+_1,\rho^+_2, \ldots \}$ in \Cref{sec:displacement_tree}. The effective bias for a particle with displacement $kL$ is given by 
\begin{equation}\label{app:rho_k}
\rho^+_k = \sum_{\nu \in \mathcal{V}_L} \dfrac{d_{\nu}-1}{d_{\nu}} \times p_k(\nu),
\end{equation}
where $p_k(\nu)$ is the probability of being at vertex $\nu$ given a displacement of $kL$. We previously assumed that the network was aperiodic when viewed as a CTMC, so that this sequence was guaranteed to converge to the unique value $\rho^+_{\infty}$ given in \cref{rho_inf}. However, if instead the network is periodic then the probability $p_k (\nu)$ converges to a periodic sequence $\{ p_{\infty,1}(\nu), \ldots , p_{\infty,m}(\nu) \}$ where $m$ is the period. Weighting these probabilities by $(d_{\nu}-1)/d_{\nu}$ we calculate the effective probabilities to move to the right along the displacement tree as
\begin{equation}
\rho^+_{\infty , j}= \sum_{\nu \in \mathcal{V}_L} \dfrac{d_{\nu}-1}{d_{\nu}} \times p_{\infty , j}(\nu),
\end{equation}
for $j \in \{1,\ldots , m\}$. We are now able to use the multiple scales approach of \Cref{PDEs} for the case of periodic networks. The periodic unit interval must be extended to have $mN+1$ lattice sites, and there will now be internal lattice sites that have asymmetric rates. The analysis proceeds identically to before, the only changes are in the size and entries of the matrices in \cref{eq:matrices}.
  


\end{document}